\documentclass[aps,prx,twocolumn,showkeyword,showpacs,superscriptaddress]{revtex4}
\usepackage{amsmath}
\usepackage[english]{babel}
\usepackage{chapterbib}
\usepackage{amssymb}
\usepackage{epsfig}
\usepackage{color}
\usepackage{graphicx,amsmath}
\graphicspath{{Fig/}}
\makeatletter
\def\@biblabel#1{#1. }
\makeatother
\begin{document}

\title{Strongly correlated Fermi systems as a new state of matter}

\author{V. R. Shaginyan}
\email{vrshag@thd.pnpi.spb.ru} \affiliation{Petersburg Nuclear
Physics Institute, NRC Kurchatov Institute, Gatchina, 188300,
Russia} \affiliation{Clark Atlanta University, Atlanta, GA
30314, USA}\author{A. Z. Msezane}\affiliation{Clark Atlanta
University, Atlanta, GA 30314, USA}\author{G. S.
Japaridze}\affiliation{Clark Atlanta University, Atlanta, GA
30314, USA} \author{K. G. Popov}\affiliation{Komi Science
Center, Ural Division, RAS, Syktyvkar, 167982, Russia}\author{V.
A. Khodel} \affiliation{Russian Research Centre Kurchatov
Institute, Moscow, 123182, Russia} \affiliation{McDonnell Center
for the Space Sciences \& Department of Physics, Washington
University, St. Louis, MO 63130, USA}

\begin{abstract}
{ The aim of this review paper is to expose a new state of
matter exhibited by strongly correlated Fermi systems
represented by various heavy-fermion (HF) metals,
two-dimensional liquids like $\rm ^3He$, compounds with quantum
spin liquids, quasicrystals, and systems with one-dimensional
quantum spin liquid. We name these various systems HF compounds,
since they exhibit the behavior typical of HF metals. In HF
compounds at zero temperature the unique phase transition,
dubbed throughout as the fermion condensation quantum phase
transition (FCQPT) can occur; this FCQPT creates flat bands
which in turn lead to the specific state, known as the fermion
condensate. Unlimited increase of the effective mass of
quasiparticles signifies FCQPT; these quasiparticles determine
the thermodynamic, transport and relaxation properties of HF
compounds. Our discussion of numerous salient experimental data
within the framework of FCQPT resolves the mystery of the new
state of matter. Thus, FCQPT and the fermion condensation can be
considered as the universal reason for the non-Fermi liquid
behavior observed in various HF compounds. We show analytically
and using arguments based completely on the experimental grounds
that these systems exhibit universal scaling behavior of their
thermodynamic, transport and relaxation properties. Therefore,
the quantum physics of different HF compounds is universal, and
emerges regardless of the microscopic structure of the
compounds. This uniform behavior allows us to view it as the
main characteristic of a new state of matter exhibited by HF
compounds.}

\end{abstract}


\maketitle

\section{Introduction}

{ Strongly correlated Fermi systems are represented by a large
variety of HF metals, insulators of new type with quantum spin
liquids, quasicrystals, two-dimensional (2D) systems and liquids
like $\rm ^3He$, and systems like $\rm Cu(C_{4}H_4N_2)(NO_3)_2$
with one-dimensional (1D) quantum spin liquid. One can hardly
expect these very different systems could exhibit a universal
behavior that would allow one to view them as a new state of
matter. We name these various systems HF compounds, for, as we
shall see, they exhibit the behavior conforming to a type of HF
metals. It is well known that three phase of matter exist:
gaseous, liquid and solid, and at high enough temperatures any
of these transforms into plasma - a system of chaotically moving
nuclei and electrons with a gas-type behavior. Surprisingly, at
the lower end of the temperature scale, close to the absolute
zero such multi-particle systems as HF compounds exhibit
universal behavior of their thermodynamic, transport and
relaxation properties, that are governed by unique quantum phase
transition called the fermion condensation quantum phase
transition (FCQPT) that creates flat bands, and leads to the
specific topological state known as the fermion condensate (FC).
In this review we show both analytically and using arguments
based entirely on the experimental grounds that all these HF
compounds exhibit the universal scaling behavior formed by
quasiparticles. This universal behavior, taking place at
relatively low temperatures and induced by FCQPT, allows us to
interpret it as the main feature of the new state of matter.
Thus, whatever mechanism drives the system to FCQPT, the system
demonstrates the universal behavior, despite numerous mechanisms
or tuning parameters present at zero temperature, such as the
pressure, number density, magnetic field, chemical doping,
frustration, geometrical frustration, etc.

As we shall see in Section \ref{EQP}, at FCQPT the effective
mass $M^*$ of quasiparticles diverges; they survive FCQPT and
define the thermodynamic, transport, and relaxation properties
of HF compounds. Thus, quasiparticles form both the main
properties and the universal scaling behavior, observed in HF
compounds. Contrary to the Landau paradigm, where quasiparticles
with approximately constant effective mass $M^*$ are
continuously linked with quasiparticles of non-interacting gas,
the effective mass of the above new quasiparticles formed at
FCQPT strongly depends on temperature $T$, magnetic field $B$,
pressure $P$, and other external parameters. As a result, we
introduce an extended quasiparticle paradigm, and show that new
quasiparticles generate the non-Fermi liquid (NFL) behavior of
the physical quantities of HF compounds, that are remarkably
different from those of ordinary solids or liquids described by
the Landau Fermi liquid (LFL) theory. Then, we briefly consider
the topological properties of FC taking place beyond FCQPT, and
how FCQPT generates the universal behavior of HF metals. We also
discuss the experimental and pure theoretical arguments of the
FC state.

The rest of the paper is organized as follows: In Sections
\ref{3He}, \ref{QSL} and \ref{QC} we examine the universal
scaling behavior of the thermodynamic, transport and relaxation
properties of HF compounds represented by 2D $^3$He, magnets of
new types with quantum spin liquid (QSL) and the recently
discovered quasicrystals, respectively, and show that HF
compounds demonstrate the new state of matter. Section \ref{TLL}
presents a perspective of materials with 1D quantum spin liquid
that can exhibit the properties of the new state of matter.
Section \ref{SUM} summaries the main results, placing the stress
on the observation that the quantum physics of different
strongly correlated Fermi systems is universal and emerges
regardless of their underlying microscopic details. This uniform
behavior, formed by flat bands, manifests the new state of
matter.}

\section{Extended quasiparticle paradigm and the scaling
behavior of HF metals}\label{EQP}

To analyze dependence of the effective mass $M^*$ on temperature
$T$, magnetic field $B$, momentum ${ p}$, number density $x$
etc., we use the Landau equation for the effective mass
\cite{landau:1957,landau:1959,lifshitzem:1980}
\begin{equation}\label{FLL} \frac{1}{M^*(T,B)} =
\frac{1}{M}+\int \frac{{\bf p}_F{\bf p_1}}{p_F^3} F({ p_F},{
p}_1,n)\frac{\partial n({ p_1},T,B)}{\partial { p_1}} \frac{d{
p}_1}{(2\pi)^3},
\end{equation}
where the Fermi-Dirac distribution reads
\begin{equation} n_{\pm}({
p},T,B)= \left\{1+\exp\left[\frac{[\varepsilon({ p},T)\pm
B\mu_B-\mu]} {T}\right]\right\}^{-1}.\label{FL4}
\end{equation}
Here, $M$ is the bare mass, $\mu$ is the chemical potential, $B$
is an external magnetic field, $n$ is quasiparticle distribution
function, and $\mu_B$ is the Bohr magneton. The term $\pm
B\mu_B$ entering the right hand side of Eq. \eqref{FL4}
describes Zeeman splitting. Equation \eqref{FLL} is exact and
can be derived within the framework of the Density Functional
Theory \cite{shaginyan:2010,amusia:2015}. This equation allows
us to calculate the behavior of $M^*$ which now becomes a
function of temperature $T$, external magnetic field $B$, number
density $x$, pressure $P$, etc.
\begin{figure} [! ht]
\begin{center}\vspace*{-0.2cm}
\includegraphics [width=0.47\textwidth]{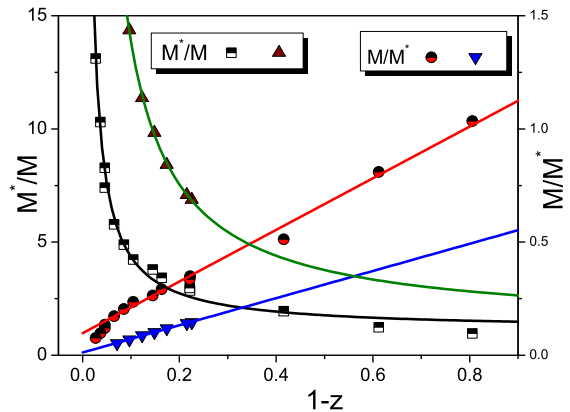}
\end{center}\vspace*{-0.5cm}
\caption {(color online) The dependence of the effective mass
$M^*(z)$ of 2D $\rm ^3He$ on dimensionless density $z=x/x_{FC}$.
Experimental data from Ref. \cite{casey:1998} are shown by
circles and squares and those from Ref. \cite{neumann:2007} are
shown by triangles. The effective mass $M^*/M$ is fitted by Eq.
\eqref{FL7}, while the reciprocal $M/M^*(z)\propto a_3\,z$,
where $a_3$ is a constant.} \label {Fig6}
\end{figure}
It is this feature of $M^*$ that forms both the scaling and the
non-Fermi liquid (NFL) behavior observed in measurements on
strongly correlated Fermi systems
\cite{khodel:1994,shaginyan:2010,shaginyan:2004:A,amusia:2015}.
In case of finite $M^*$ and at $T=0$ the distribution function
$n({ p},T=0)$ becomes the step function $\theta (p_F-p)$, as it
follows from Eq. \eqref{FL4}, and Eq. (\ref{FLL}) yields the
well-known result
\begin{equation}\label{M*}
\frac{M^*}{M}=\frac{1}{1-F^1/3},
\end{equation}
where $F^1=N_0f^1$, $f^1(p_F,p_F)$ is the $p$-wave component of
the Landau interaction, $p_F$ is the Fermi momentum, and
$N_0=Mp_F/(2\pi^2)$ is the density of states (DOS) of a free
Fermi gas. Because the number density $x=p_F^3/3\pi^2$, the
Landau interaction can be written as $F^1(p_F,p_F)=F^1(x)$
\cite{shaginyan:2010}. At a certain critical point $x=x_{FC}$,
with $x\to x_{FC}$ from below, the denominator $(1-F^1(x)/3)$
tends to zero
\begin{equation}\label{MM*}
(1-F^1(x)/3)\propto(x_{FC}-x)+a(x_{FC}-x)^2 + ...\to 0,
\end{equation}
and we find that
\begin{equation}
\frac{M^*(x)}{M}\simeq
a_1+\frac{a_2}{1-x/x_{FC}}=a_1+\frac{a_2}{1-z},
\label{FL7}\end{equation} where $x/x_{FC}=z$, $a_1$, and $a_2$
are constants, and $M^*(x\to x_{FC})\to\infty$. As a result, at
$T\to0$ and $x\to x_{FC}$ the system undergoes a quantum phase
transition \cite{shaginyan:2010,amusia:2015}, represented by
FCQPT. We note that the divergence of the effective mass given
by Eq. \eqref{FL7} preserves the Pomeranchuk stability
conditions, for $F^1$ is positive, rather than negative
\cite{khodel:1994,pomeranchuk:1958:A,nozieres:1992}. The
divergence of the effective mass $M^*(x)$, given by (\ref{FL7})
and observed in measurements on two-dimensional $^3$He
\cite{casey:1998,neumann:2007}, is illustrated in Fig.
\ref{Fig6}. It is seen that the calculations are in good
agreement with the data \cite{shaginyan:2008}.

\subsection{Topological properties of systems with fermion
condensate}\label{FC} 
\begin{figure} [! ht]
\begin{center}
\vspace*{-0.2cm}
\includegraphics [width=0.47\textwidth]{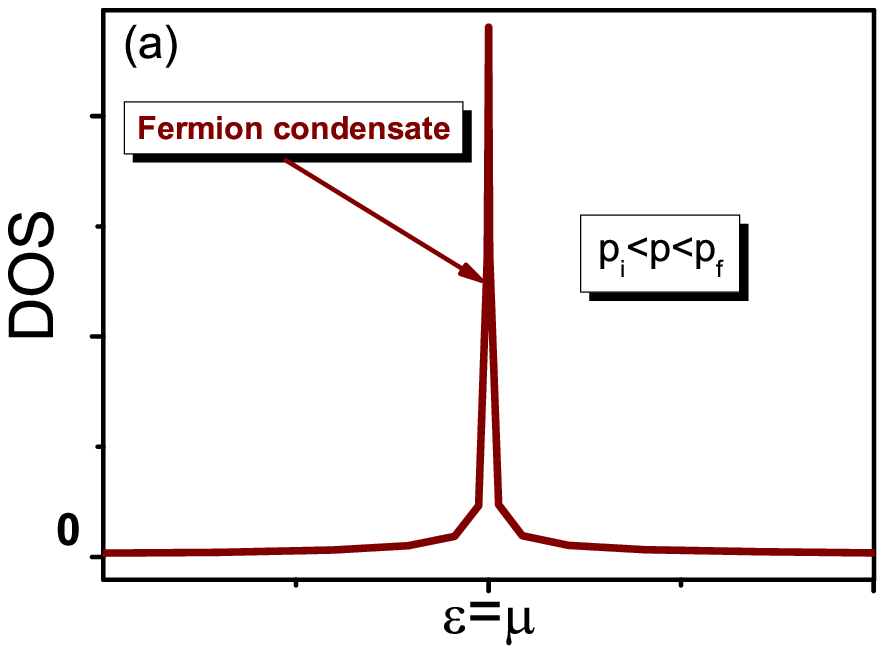}
\includegraphics [width=0.47\textwidth]{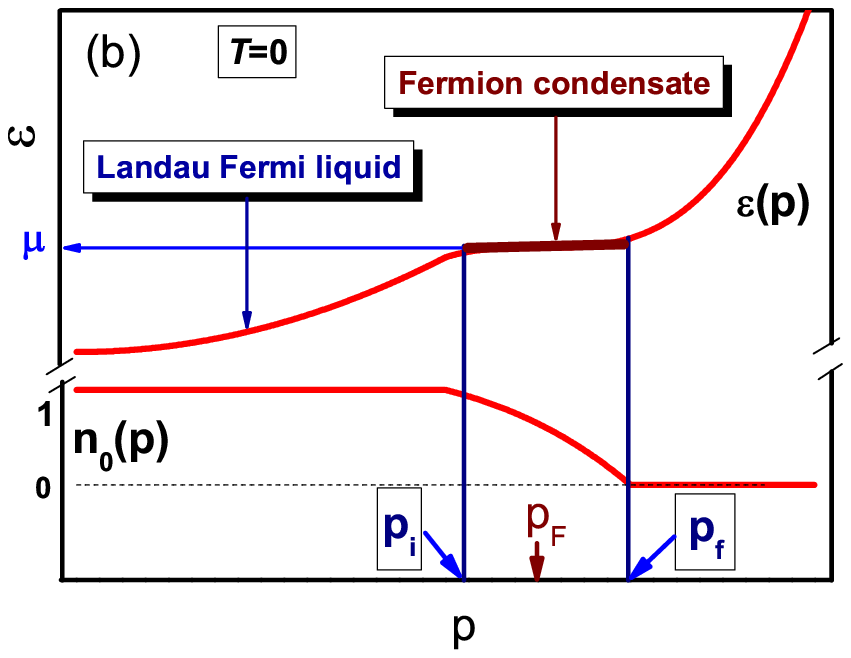}
\end{center}
\vspace*{-0.7cm} \caption {(color online) Fermion condensate.
Panel  (a): Schematic plot of the density of states (DOS) at low
temperatures of quasiparticles versus $\varepsilon$ at the
momentum $p_i<p<p_f$ of a Fermi liquid with FC. Panel (b):
Schematic plot of two-component Fermi liquid at $T=0$ with FC.
The system is separated into two parts shown by the arrows: The
first part is a Landau Fermi liquid with the quasiparticle
distribution function $n_0(p<p_i)=1$, and $n_0(p>p_f)=0$; The
second one is FC with $0<n_0(p_i<p<p_f)<1$ and the
single-particle spectrum $\varepsilon (p_i<p<p_f)=\mu$. The
Fermi momentum $p_F$ satisfies the condition $p_i<p_F<p_f$.}
\label{fig0}
\end{figure}

The quasiparticle distribution function
$n( p)$ of Fermi systems with FC is determined by the usual
equation for a minimum of the Landau functional $E[n( p)]$. In
contrast to common functionals of the number density $x$
\cite{hohenberg:1964,kohn:1965}, the Landau functional of the
ground state energy $E$ becomes functional of the occupation
numbers $n$ \cite{khodel:1994,shaginyan:2010,amusia:2015}. In
case of homogeneous system a common functional becomes a
function of $x=\sum_p n({ p})$, while the Landau functional
$E=E[n({ p})]$ \cite{khodel:1994,shaginyan:2010,khodel:1990}
satisfies
\begin{equation} \frac{\delta E[n({ p})]}{\delta
n({ p})}=\varepsilon({ p})=\mu;\,\, {\rm when\,} 0<n({ p})<1.
\label{FCM}\end{equation} The minimum of the functional $E$ has
to be found from Eq. \eqref{FCM}. { While in the case of Bose
system equation $\delta E/\delta n(p)=\mu$ is well established
and understood, in case of Fermi system this equation, generally
speaking , is not correct. It is the above constraint, $0\leq
n({ p})\leq 1$, dictated by the Fermi distribution function that
makes Eq. \eqref{FCM} applicable for Fermi systems. As we have
seen above, in the case of LFL liquid at $T=0$ the distribution
function is represented by the step function: $n(p<p_F)=1$ and
$n(p>p_F)=0$; while in the case of FC in some region of momenta,
$p_i<p<p_f$, the distribution function becomes a smooth function
at the Fermi surface, $0<n({ p})<1$. Because of the above
constraint, in the region $p_i<p<p_f$, Fermi quasiparticles can
behave as Bose one, occupying the same energy level
$\varepsilon=\mu$, and Eq. \eqref{FCM} yields the quasiparticle
distribution function $n_0({ p})$ that minimizes the
ground-state energy $E$. Thus, at $T=0$ the system undergoes a
topological phase transition, for the Fermi surface at $p=p_F$
transforms into the Fermi volume for $p_i\leq p\leq p_f$
suggesting that at $T=0$ the band is absolutely ``flat'' within
this interval, giving rise to the spiky DOS as seen from the
panel (a) of Fig. \ref{fig0}. A possible solution $n_0({ p})$ of
Eq. (\ref{FCM}) and the corresponding single-particle spectrum
$\varepsilon({ p})$ are depicted in Fig. \ref{fig0}, panel (b).
As seen from the panel (b), $n_0({ p})$ differs from the step
function in the interval $p_i<p<p_f$, where $0<n_0({ p})<1$, and
coincides with the step function outside this interval. The
existence of such flat bands, formed by inter-particle
interaction, was predicted for the first time in Ref.
\cite{khodel:1990}. Quasiparticles with momenta within the
interval $(p_i<p<p_f)$ have the same single-particle energies
equal to the chemical potential $\mu$ and form FC, while the
distribution $n_0({ p})$ describes the new state of the Fermi
liquid with FC, and the Fermi system is split up into two parts:
LFL and the FC part, as shown in Fig. \ref{fig0}, panel (b)
\cite{khodel:1990,khodel:1994,volovik:1991,heikkila:2011,volovik:2015,heikkila:2015,shaginyan:2010,amusia:2015}.}

The theory of fermion condensation permits the construction of
new class of strongly correlated Fermi liquids with the fermion
condensate
\cite{khodel:1990,khodel:1994,volovik:1991,heikkila:2011,volovik:2015,shaginyan:2010,heikkila:2015}.
In that case the quasiparticle system is composed of two parts:
One of them is represented by FC located at the chemical
potential $\mu$, and giving rise to the spiky DOS, like that
with $p=0$ of Bose systems, as seen from Fig. \ref{fig0}, panel
(a), that shows the DOS of a Fermi liquid with FC located at the
momenta $p_i<p<p_f$ and energy $\varepsilon=\mu$. Contrary to
the condensate of a Bose system occupying the $p=0$ state,
quasiparticles of FC with the energy $\varepsilon=\mu$ must be
spread out over the interval $p_i<p<p_f$.
Contrary to the Landau, marginal, superconducting LFL, or
Tomonoga-Luttinger (marginal) Fermi liquids whose Green's
functions exhibit the same topological behavior, in systems with
FC, where the Fermi surface spreads into the Fermi volume, the
Green's function belongs to a different topological class. The
topological class of the Fermi liquid is characterized by the
topological invariant
\cite{volovik:1991,heikkila:2011,volovik:2015, heikkila:2015}
\begin{equation} N=tr\oint_C\frac{dl}{2\pi i}G(i\omega,{ p})
\partial_lG^{-1}(i\omega,{ p})\label{FLVOL},\end{equation}
where ``tr'' denotes the trace over the spin indices of the
Green's function and the integral is calculated along an
arbitrary contour $C$ encircling the singularity of the Green's
function. The invariant $N$ in \eqref{FLVOL} assumes integer
values even when the singularity is not a simple pole, cannot
vary continuously, and is conserved in a transition from the
Landau Fermi liquid to marginal liquids and under small
perturbations of the Green's function. As shown by Volovik
\cite{volovik:1991}, the situation is quite different for
systems with FC, where the invariant $N$ becomes a half-integer
and the system with FC transforms into an entirely new class of
Fermi liquids with its own topological structure, preserving the
FC state, and forming the new state of matter demonstrated by
many HF compounds
\cite{amusia:2015,amusia:2013,volovik:1991,heikkila:2011,heikkila:2015,volovik:2015,yudin:2014}.

In contrast to Bose liquid, whose entropy $S\to0$ at temperature
$T\to0$, a Fermi liquid with FC possesses finite entropy $S_0$
at zero temperature \cite{shaginyan:2010,zverev:1997:A}. Indeed,
as seen from Fig. \ref{fig0}, panel (b), at $T=0$, the ground
state of a system with a flat band is degenerate, and the
occupation numbers $n_0({ p})$ of single-particle states
belonging to the flat band are continuous functions of momentum
${ p}$, in contrast to discrete standard LFL values 0 and 1.
Such behavior of the occupation numbers leads to a
$T$-independent entropy term $S_0=S(T\to0,n=n_0)$ with the
entropy given by
\begin{equation}
S(n)=-\sum_{\,{ p}} [n({ p})\ln n({ p})+(1-n({ p}))\ln(1- n({
p}))].\label{S0}
\end{equation}
Since the state of a system with FC is highly degenerate, FC
triggers phase transitions that could lift the degeneracy of the
spectrum and vanish $S_0$ in accordance with the Nernst theorem.
For instance, FC can excite the formation of spin density waves,
antiferromagnetic state, ferromagnetic state, the
superconducting state etc., thus strongly stimulating the
competition between phase transitions eliminating the
degeneracy. Contrary to LFL, where the entropy vanishes at zero
temperature, the finite entropy $S_0$, characteristic of Fermi
liquid with FC, causes the emergence of diversity of states.
This observation is in accordance with the experimental phase
diagrams \cite{amusia:2015,lohneysen:1996}. { Since at $T=0$ the
entropy of ordered states is zero, we conclude that the entropy
is discontinuous at the phase transition point, with its
discontinuity $\delta S=S_0$. Thus, the entropy suddenly
vanishes, with the system undergoing the first-order transition
near which the critical quantum and thermal fluctuations are
suppressed and the quasiparticles are well-defined excitations
\cite{shaginyan:2010,af2010}. As a result, we conclude that the
main contribution to the transport and thermodynamic properties
comes from quasiparticles rather than the various low energy
boson excitations. We note that the existence of FC has been
convincingly demonstrated by purely theoretical arguments, see
e.g.
\cite{lids,katanin,lee,shaginyan:2010,yudin:2014,volovik:2015,amusia:2015},
and by experimental facts, see e.g.
\cite{shaginyan:2010,shash2014,amusia:2015,shash2016}}.

There are different kinds of instabilities of normal Fermi
liquids connected with several perturbations of initial
quasiparticle spectrum $\varepsilon({ p})$ and occupation
numbers $n({ p})$, associated with the emergence of a
multi-connected Fermi surface, see e.g.
\cite{amusia:2015,volovik:1991,heikkila:2011,
volovik:2015,heikkila:2015,khodel:2008}. Depending on the
parameters and analytical properties of the Landau interaction,
such instabilities lead to several possible types of
restructuring of the initial Fermi liquid ground state. This
restructuring generates topologically distinct phases. One of
them is the FC, and the other belongs to a class of topological
phase transitions, where the sequence of rectangles $n(p)=0$ and
$n(p)=1$ is realized at $T=0$, see e.g. \cite{art98}. In fact,
at elevated temperatures the systems located at these
transitions exhibit behavior typical to those located at FCQPT.
Therefore, we do not consider the specific properties of these
topological transitions, but focus on the behavior of systems
located near FCQPT.

\subsection{Scaling behavior of HF metals} \label{hf}

It is instructive to briefly explore the behavior of $M^*$ in
order to capture its universal behavior at FCQPT. Let us write
the quasiparticle distribution function as $n_1({ p})=n({
p},T,B)-n({ p})$, with $n({ p})$ being the step function, Eq.
\eqref{FLL} then becomes
\begin{equation}
\frac{1}{M^*(T,B)}=\frac{1}{M^*}+\int\frac{{ p}_F{
p_1}}{p_F^3}F({ p_F},{ p}_1)\frac{\partial n_1(p_1,T,B)}
{\partial p_1}\frac{d{ p}_1}{(2\pi)^3}. \label{LF1}
\end{equation}

\begin{figure} [! ht]
\begin{center}
\vspace*{-0.2cm}
\includegraphics [width=0.47\textwidth]{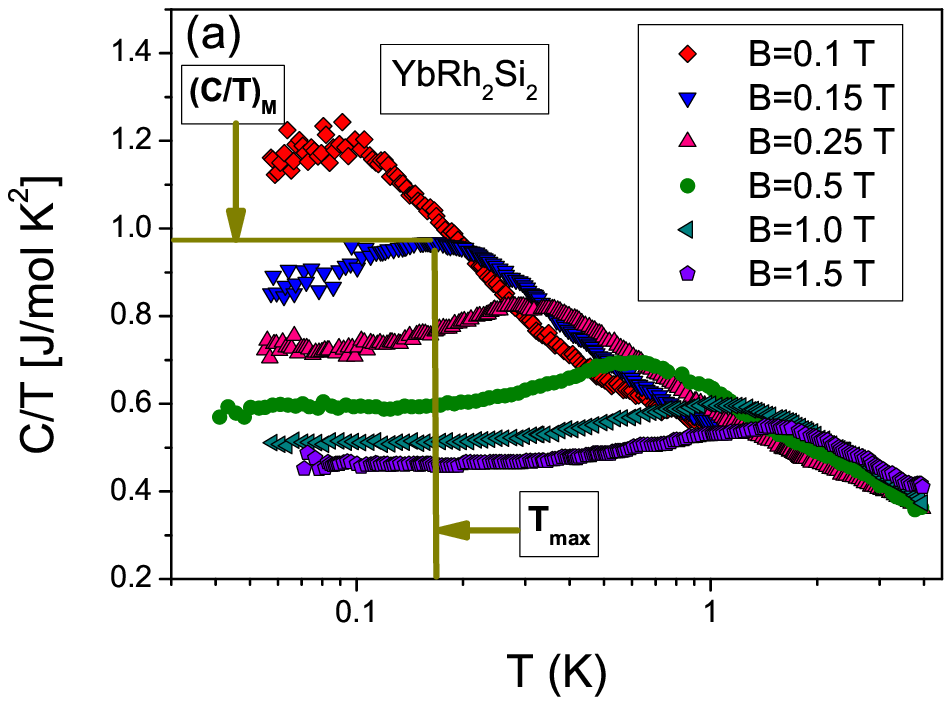}
\includegraphics [width=0.47\textwidth]{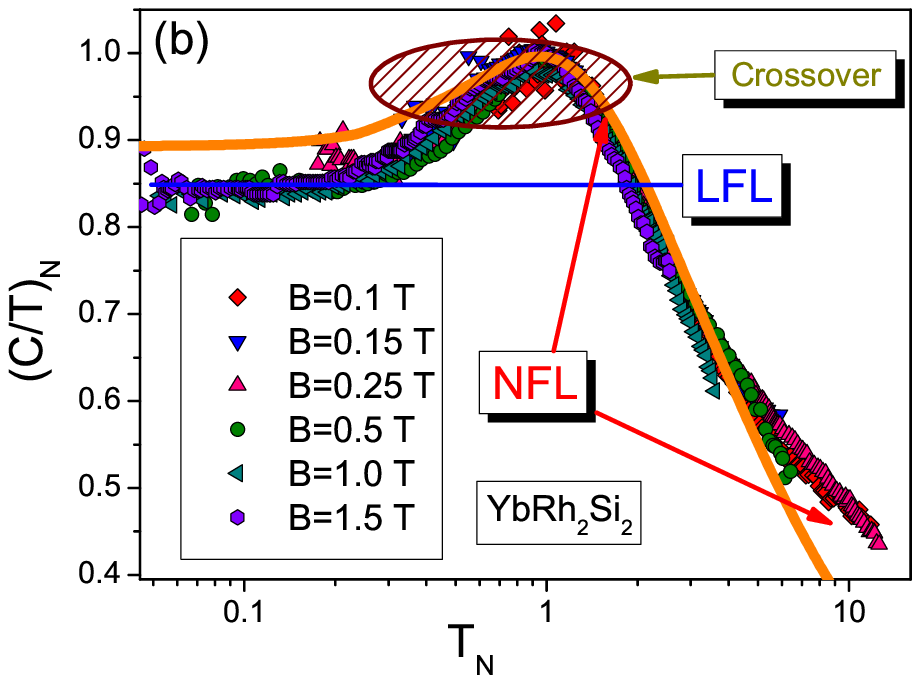}
\end{center}
\vspace*{-0.5cm} \caption{(color online) Scaling behavior of HF
metals. Panel (a): Electronic specific heat of $\rm YbRh_2Si_2$,
$C/T$, versus temperature $T$ as a function of magnetic field
$B$ \cite{oeschler:2008} shown in the legend. The illustrative
values of $(C/T)_M\propto M^*_M$ and $T_{M}$ at $B=0.15$ T are
shown. Panel (b): The normalized effective mass $M^*_N$ versus
normalized temperature $T_N$. $M^*_N$ is extracted from the
measurements of the specific heat $C/T$ on $\rm YbRh_2Si_2$
\cite{oeschler:2008}, displayed in the panel (a). The constant
effective mass inherent in a normal Landau Fermi liquid is
presented by the solid line. The schematic crossover region is
indicated by the arrow and the NFL behavior is indicated by two
arrows. Our calculation based on Eq. \eqref{FLL} is displayed by
the solid curve.}\label{FIG2}
\end{figure}

{ We now qualitatively analyze the solutions of Eq. \eqref{LF1}
at $T=0$. Application of magnetic field leads to Zeeman
splitting of the Fermi surface, and the distance $\delta p$
between the Fermi surfaces with spin up and spin down becomes
$\delta
p=p_F^{\uparrow}-p_F^{\downarrow}\sim\mu_{B}BM^*(B)/p_F$. We
note that the second term on the right-hand side of Eq.
\eqref{LF1} is proportional to $(\delta
p)^2\propto(\mu_{B}BM^*(B)/p_F)^2$, and therefore Eq.
\eqref{LF1} reduces to \cite{shaginyan:2010}
\begin{equation}
\frac{M}{M^*(B)}=\frac{M}{M^*(x)}+c\frac{(\mu_{B}BM^*(B))^2}
{p_F^4},\label{HC4}
\end{equation}
where $c$ is a constant. In the same way, one can calculate the
change of the effective mass due to the variation of $T$
\cite{shaginyan:2010}. For normal metals, where the electron
liquid behaves like LFL with the effective mass of several bare
electron masses $M^*/M\sim1$, at temperatures even near 1000 K,
the second term on the right hand side of Eq. \eqref{LF1} is of
the order of $T^2/\mu^2$ and is much smaller than the first
term. The same is true, as can be verified, when a magnetic
field of reasonable strength of $B\sim100$ T is applied. Thus,
the system behaves like LFL with the effective mass that is
actually independent of the temperature or magnetic field, while
the resistivity $\rho(T)\propto T^2$. This means that the
correction to the effective mass determined by the second term
on the right-hand side of Eq. \eqref{HC4} is small; thus the
effective mass is approximatively constant under the influence
of external parameters of reasonable strength. We recall that in
the case of common metals $\mu \sim 1$ eV, therefore for
reasonable temperatures $T$ and magnetic fields $B$ one obtain
$T/\mu\ll 1$ and $\mu_BB/\mu\ll 1$. As a result, the integral on
the right hand side of Eq. \eqref{FLL} represents a small
correction to $M^*$ as a function of $T$ and $B$, provided that
$M/M^*\sim 1$. At FCQPT, e.g. as soon as $x\to x_{FC}$, the
effective mass $M^*(x)$ diverges and $M/M^*(x)\to0$. In that
case, the first term on the right hand side of Eq. \eqref{HC4}
vanishes; the second term becomes principal, and determines
$M^*$ as a function of $B$. In the same way, Eq. \eqref{LF1}
becomes homogeneous, and determines $M^*(T,B)$ as a universal
function of temperature, magnetic field and other tuning
parameters that drive the system to FCQPT. As it is seen from
Fig. \ref{FIG2} (a) and (b), our observations are in accordance
with the experimental facts: Under the application of low
magnetic fields of $0.1$ T the HF metal $\rm YbRh_2Si_2$
exhibits LFL behavior at $T\simeq 0.1$ K, while the effective
mass $M^*$ is strongly dependent on both $B$ and $T$.} The only
role of the Landau interaction $F$ is to drive the system to
FCQPT and the solutions of Eq. \eqref{LF1} are represented by
some universal function of the variables $T$, $B$, $x$. In that
case $M^*$ strongly depends on the same variables. In contrast
to the Landau quasiparticle paradigm that assumes approximate
constancy of the effective mass, the extended quasiparticle
paradigm is to be introduced. The main point here is that the
well-defined quasiparticles determine as before the
thermodynamic, relaxation and transport properties of strongly
correlated Fermi-systems, while $M^*$ becomes a function of $T$,
$B$, $x$, etc. Moreover, the effective mass can be a divergent
function of $T$, $M^*(T\to0)\to\infty$ \cite{shaginyan:2010}.
Thus, we have to introduce the modified and extended
quasiparticles paradigm, that permits to naturally describe the
basic properties and the scaling behavior of both the $M^*$ and
HF compounds. { The essence of the paradigm is that in spite of
the altering of Fermi surface topology the substance undergoes
at FCQPT, the Landau quasiparticles survive but completely
change their properties
\cite{shaginyan:2010,amusia:2015,khod2scen,khodel:2008}.}

A deeper insight into the behavior of $M^*(T,B)$ can be achieved
by using some "internal" scales. Namely, near FCQPT the
solutions of Eq. \eqref{LF1} exhibit a behavior such that
$M^*(T,B)$ reaches its maximum value $M^*_M$ at some temperature
$T_{M}\propto B$ \cite{shaginyan:2010}. It is convenient to
introduce the internal scales $M^*_M$ and $T_{M}$ to measure the
effective mass and temperature. We rescale $M^*$ and $T$ by
$M^*_M$ and $T_{M}$. This generates the normalized dimensionless
effective mass $M^*_N=M^*/M^*_M$ and the corresponding
normalized dimensionless temperature $T_N=T/T_{M}$.
As an illustration to the above consideration, we analyze the
specific heat $C/T\propto M^*$ of the HF metal $\rm YbRh_2Si_2$
\cite{oeschler:2008}. When the  magnetic field $B$ is applied,
the specific heat exhibits a behavior that is described by a
function of both $T$ and $B$. As seen from the panel (a) of Fig.
\ref{FIG2}, a maximum structure $(C/T)_M$ in $C/T\propto M^*_M$
at temperature $T_{M}$ appears when $B$ is applied. $T_{M}$
shifts to higher $T$ and $C/T\propto M^*_M$ diminishes as $B$ is
increased. The value of $C/T$ reaches its maximum at lower
temperatures, which decreases at elevated magnetic field. To
obtain the normalized effective mass $M^*_N$, we use $(C/T)_M$
and $T_{M}$ as "internal" scales: the maximum structure
$(C/T)_M$ was used to normalize $C/T$, and $T$ was normalized by
$T_{M}$. In panel (b) of Fig. \ref{FIG2}, the obtained
$M^*_N=(C/T)/(C/T)_M=M^*/M^*_M$, as a function of normalized
temperature $T_N=T/T_{M}$, is shown. It is seen that the LFL
state with $M^*=const$ and NFL one are separated by the
crossover at which $M^*_N$ reaches its maximum value. The panel
(b) of Fig. \ref{FIG2} reveals the scaling behavior of the
normalized experimental curves: the curves at different magnetic
fields $B$ merge into a single one in terms of the normalized
variable $T/T_M$. Our calculations of the normalized effective
mass $M^*_N(T_N)$, shown by the solid line and based on Eq.
\eqref{LF1}, are in good agreement with the data
\cite{shaginyan:2004:A,shaginyan:2010,amusia:2015,shaginyan:2009}.
Near FCQPT the normalized solution of Eq. \eqref{FLL}
$M^*_N(T_N)$ can be approximated well by a simple universal
interpolating function \cite{shaginyan:2010}. The interpolation
occurs between the LFL and NFL regimes and represents the
universal scaling behavior of $M^*_N$
\cite{shaginyan:2010,shaginyan:2009}
\begin{equation}M^*_N(T_N)\approx c_0\frac{1+c_1T_N^2}{1+c_2T_N^{n}}.
\label{UN2}
\end{equation}
Here, $c_0=(1+c_2)/(1+c_1)$, $c_1$, $c_2$ are fitting
parameters; and the exponent $n=8/3$ if the Landau interaction
is an analytical function, otherwise $n=5/2$ \cite{shaginyan:2010}. It follows
from Eq.~\eqref{LF1} that
\begin{equation}
\label{TMB} T_M\simeq a_1\mu_BB,
\end{equation}
where $a_1$ is a dimensionless number. { Equation \eqref{TMB} is
in good agreement with experimental facts
\cite{brando:2013,shaginyan:2014}, while possible corrections to
Eq. \eqref{TMB}, taking place near the corresponding phase
transitions, are discussed in Refs.
\cite{amusia:2015,shaginyan:2014}. Here we do not discuss these
correction, for the main goal of our review is to expose the new
state of matter exhibited by strongly correlated Fermi systems}.
Note that the effective mass $M^*$ defines the thermodynamic
properties of HF compounds, therefore $M^*(T)\propto
C(T)/T\propto S(T)/T\propto M_0(T)\propto\chi(T)$ where $C(T)$
is the specific heat, $S(T)$
--- entropy, $M_0(T)$ --- magnetization and $\chi(T)$
--- AC magnetic susceptibility. For the normalized values
we have
\begin{equation}
\label{NORM} M^*_N=(C/T)_N=(S/T)_N=(M_0)_N=\chi_N.
\end{equation}

\begin{figure} [! ht]
\begin{center}\vspace*{-0.2cm}
\includegraphics [width=0.46\textwidth]{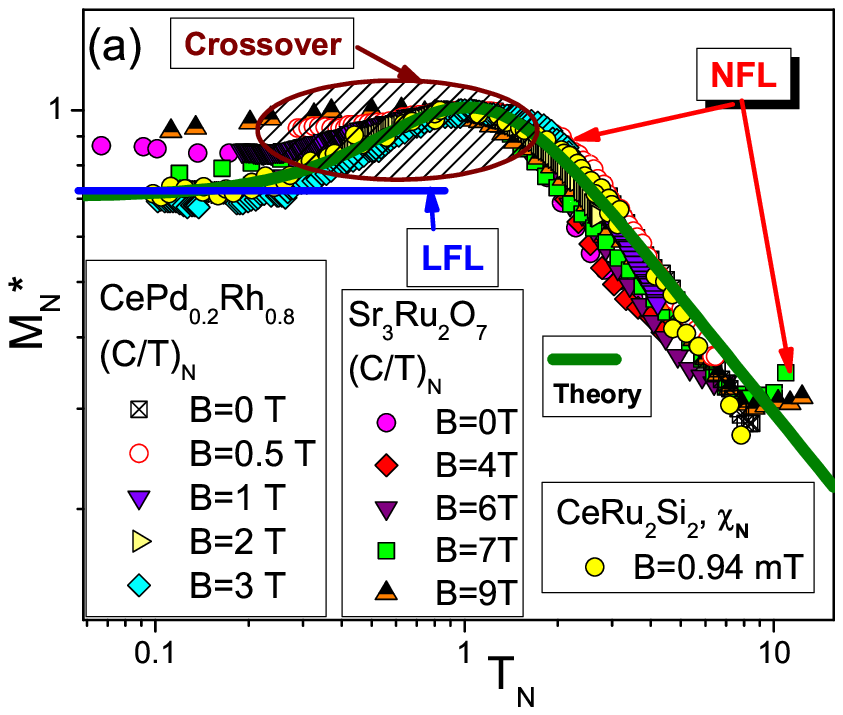}
\includegraphics [width=0.50\textwidth]{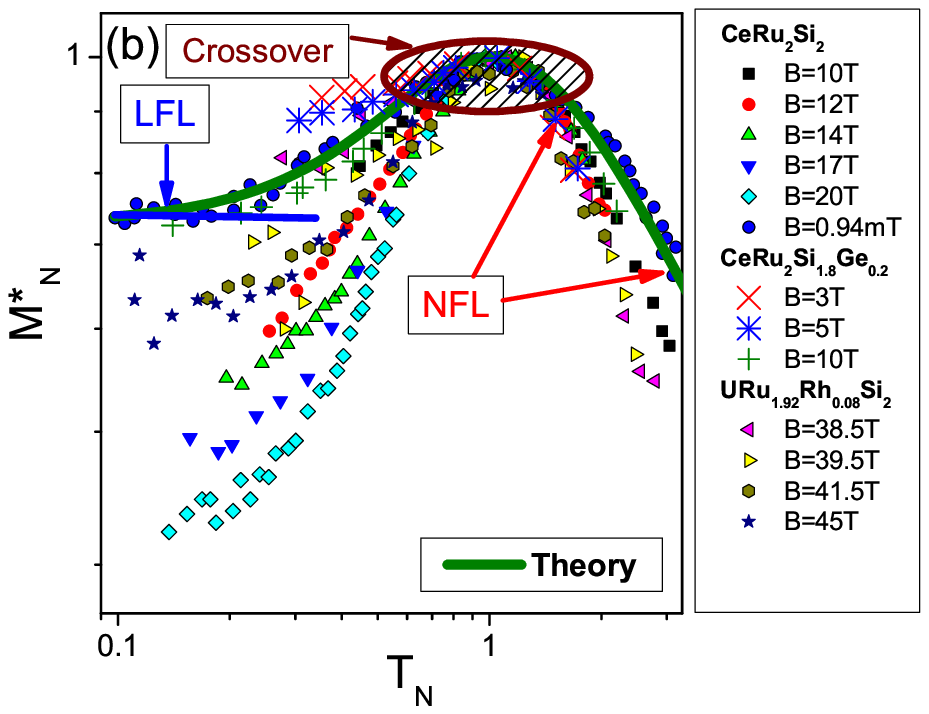}
\end{center}
\vspace*{-0.5cm} \caption{(color online) The universal scaling
behavior of the normalized effective mass $M^*_N$ versus $T_N$.
Panel (a): $M^*_N$ is extracted from the measurements of $\chi$
and $C/T$ (in magnetic fields $B$ shown in the legends) on $\rm
CeRu_2Si_2$ \cite{takahashi:2003}, $\rm CePd_{1-x}Rh_x$ with
$x=0.80$ \cite{oeschler:2008}, and $\rm Sr_3Ru_2O_7$
\cite{rost:2011}. The LFL and NFL regimes (latter having
$M^*_N\propto T_N^{-2/3}$) are shown by the arrows and straight
lines. The transition regime is depicted by the shaded area. The
solid curve represents our calculated universal behavior of
$M^*_N(T_N)$. Panel (b): The normalized effective mass versus
the normalized temperature at different magnetic fields $B$,
shown in the legend. $M^*_N(T_N)$ is extracted from measurements
of $C/T$ collected on $\rm URu_{1.92}Rh_{0.08}Si_2$, $\rm
CeRu_2Si_2$ and $\rm CeRu_2Si_{1.8}Ge_{0.2}$ at their
metamagnetic phase transitions\cite{silhanek:2005,kim:1990}. The
solid curve gives the universal behavior of $M^*_N$, Eq.
(\ref{UN2}).}\label{metaSrRuO}
\end{figure}
It is seen from Eq. \eqref{UN2} and Figs. \ref{FIG2} and
\ref{metaSrRuO}, that at elevated temperatures the considered HF
compounds exhibit the NFL behavior, $M^*(T)\propto T^{-2/3}$.
NFL behavior manifests itself in the power-law behavior of the
physical quantities of strongly correlated Fermi systems located
close to their QCPs \cite{takahashi:2003,rost:2011}, including
QCPs related to metamagnetic phase transitions
\cite{silhanek:2005,kim:1990}, with exponents different from
those of ordinary Fermi liquids. As seen from Fig.
\ref{metaSrRuO}, panels (a) and (b), $M^*_N$ in different HF
metals is the same, both in the high and low magnetic field.
This observation is of utmost importance since allows us to
verify the universal behavior in HF metals, when quite different
magnetic fields are applied
\cite{shaginyan:2010,amusia:2015,shaginyan:2011:C}. { Relatively
small values of $M^*_N$ observed in $\rm
URu_{1.92}Rh_{0.08}Si_2$ and $\rm CeRu_2Si_2$ at the high fields
and small temperatures can be explained by taking into account
that the narrow band is completely polarized. As a result, at
low temperatures the summation over up and down spin projections
reduces to a single direction producing the coefficient 1/2 in
front of $M^*_N$, thus violating the scaling at low temperatures
$T_N\leq1$. At high temperatures the polarization vanishes and
the summation is restored. As seen from Fig. \ref{metaSrRuO},
panel (b), these observations are consistent with the
experimental data collected in measurements under the
application of high magnetic fields $B$ on the archetypical HF
metal $\rm YbRh_2Si_2$ \cite{gegenwart:2006:A} the theoretical
analysis of which is in good agreement with the experimental
facts \cite{shaginyan:2011:C}. Thus, we conclude that different
HF metals exhibit the same LFL, crossover and NFL behavior in
strong magnetic fields that is in agreement with the notion of
the new state of matter.}

It is a common belief that the main output of theory is to
explain the exponents that describe the NFL behavior, e.g.
$C/T\propto T^q$,  which are at least dependent on the magnetic
character of QCP, dimensionality of the system, and employed
scenario \cite{lohneysen:2007}. On contrary, the NFL behavior
cannot be captured by these exponents as seen from Figs.
\ref{FIG2} and \ref{metaSrRuO}. Indeed, the specific heat $C/T$
exhibits a behavior that has to be described as a function of
both temperature $T$ and magnetic field $B$ rather than by a
single exponent. One can see that at low temperatures $C/T$
demonstrates the LFL behavior which is changed by the crossover
regime at which $C/T$ reaches its maximum and finally $C/T$
decays into NFL behavior as a function of $T$ at fixed $B$. It
is clearly seen from Figs. \ref{FIG2} and \ref{metaSrRuO} that,
in particularly in the transition regime, these exponents may
have little physical significance. { Thus, it is the
quasiparticles of the extended paradigm that form the universal
scaling behavior, and reproduce the salient experimental data
\cite{shaginyan:2010,amusia:2015}.

We note that there have been developed the Kondo breakdown
theory, see e.g. \cite{coleman2001,coleman2003,coleman2004}, the
two-fluid description of heavy electron leading to quantum
critical behavior, see e. g.
\cite{yang:2008n,yang:2008,yang:2014}, and critical
quasiparticle theory, see e. g.
\cite{abrahams2011,abrahams:2012}. These perspective theories
consider hybridization between conduction electrons and local
magnetic moments, e. g. $f$ orbitals, that produces a heavy
electron liquid with an associated mass enhancement. As a
result, one observes the itinerant heavy electron liquid in
materials that contain a Kondo lattice of localized $f$
electrons coupled to background conduction electrons. Thus, the
Kondo effect in many materials develops quantum criticality that
emerges from a competition between local moment magnetism and
the conduction electron screening of the local moments
\cite{s2011,s2015}.

Interesting example of quantum criticality is represented by the
heavy-fermion superconductor $\rm \beta-YbAlB_4$: It is
suggested that $\rm \beta-YbAlB_4$ exhibits strange metallic (or
NFL) behavior across an extensive pressure regime, distinctly
separated from a high-pressure magnetic quantum phase transition
by LFL phase \cite{s2011,s2015}. Moreover, the superconductor
$\rm \beta-YbAlB_4$ demonstrates the robustness of the NFL
behavior of the thermodynamic properties and of the anomalous
$T^{3/2}$ temperature dependence of the electrical resistivity
under applied pressure in zero magnetic field $B$; such a
behavior is at variance with the fragility of the NFL phase
under application of tiny magnetic field \cite{s2011,s2015},
strongly resembling the corresponding behavior observed in the
quasicrystal $\rm Au_{51}Al_{34}Yb_{15}$
\cite{deguchi:2012,shaginyan:2013:A}, as we shall see in Section
\ref{QC}. A consistent topological basis for this behavior
observed in both the HF metal $\rm \beta-YbAlB_4$ and the
quasicrystal $\rm Au_{51}Al_{34}Yb_{15}$, as well as the
empirical scaling laws, may be found within the
fermion-condensation theory \cite{shaginyan:2013:A,shag:2016}.}
Then, as we shall see, the FC theory allows us to reveal and
explain the universal behavior of two-dimensional liquids like
$\rm ^3He$, Section \ref{3He}, compounds with quantum spin
liquids, Section \ref{QSL}, quasicrystals, Section \ref{QC}, and
systems with 1D quantum spin liquid, Section \ref{TLL}.
Therefore, in our short review devoted to revealing the new
state of matter, we employ the FC theory.

Several remarks concerning the applicability of Eqs. \eqref{FLL}
and \eqref{UN2} to systems with violated translational
invariance are in order. We study the universal behavior of HF
metals, quantum spin liquids, and quasicrystals at low
temperatures using the model of a homogeneous HF liquid
\cite{shaginyan:2010,amusia:2015}. The model is applicable
because we consider the scaling behavior exhibited by the
thermodynamic properties of these materials at low temperatures,
a behavior related to the scaling of quantities such as the
effective mass $M^*$, the heat capacity $C/T\propto M^*$, the
magnetic susceptibility $\chi\propto M^*$, etc. The behavior of
$M^*_N(T_N)$, that is used to describe quantities mentioned
above, is determined by momentum transfers that are small
compared to momenta of the order of the reciprocal lattice
length. The high momentum contributions can therefore be ignored
by substituting the lattice with the jelly model. The values of
the scales, like the maximum $M^*_M(B_0)$ of the effective mass
measured at some field $B=B_0$ and $T_M$ at which $M^*_M$
emerges, are determined by a wide range of momenta. Thus, these
scales are controlled by the specific properties of the system
under consideration, while the scaled thermodynamic properties
of different strongly correlated Fermi systems can be described
by universal function \eqref{UN2} determining $M^*_N(T_N)$. It
is seen from Figs. \ref{FIG2} and \ref{metaSrRuO}, and
demonstrated by numerous experimental data collected on HF
compounds that this observation is in a good agreement with
experimental data \cite{shaginyan:2010,amusia:2015}, and allows
one to view the universal scaling behavior as a manifestation of
the new state of matter exhibited by HF compounds.

\subsection{Universal behavior of the thermopower $S_T$ of heavy-fermion metals}

In this section, we demonstrate that the thermopower $S_T$ of
such different heavy-fermion (HF) compounds as
$\mathrm{YbRh_2Si_2}$, $\beta$-${\rm YbAlB_4}$ and the strongly
correlated layered cobalt oxide $\rm
[BiBa_{0.66}K_{0.36}O_{2}]CoO_{2}$ also exhibits the universal
scaling behavior.

\begin{figure}[!ht]
\begin{center}
\includegraphics [width=0.47\textwidth]{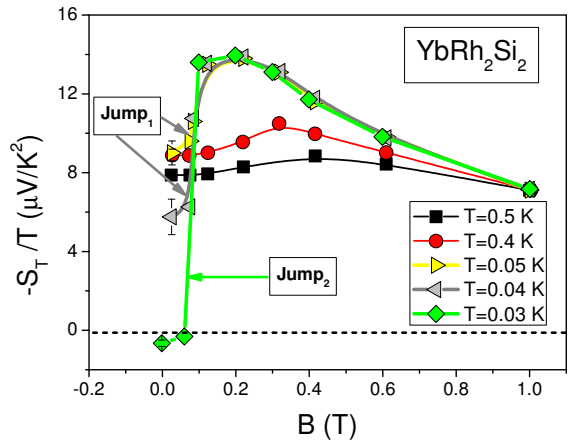}
\end{center}
\caption{(color online) Thermopower isotherm $-S_T(B)/T$ for
different temperatures shown in the legend \cite{TP,TPJ}. The
labels $\rm Jump_1$ and $\rm Jump_2$ represent the first and
second downward jumps in $-S_T(B)/T$ shown by the arrows. The
solid lines are guides to the eye.} \label{figYb}
\end{figure}
A study of the thermoelectric power $S_T$ may deliver new
insight into the nature of quantum phase transition that defines
the features of new state of matter. For example, one may
reasonably propose that the thermoelectric power $S_T$
distinguishes between two competing scenarios for quantum phase
transitions in heavy fermions, namely the spin-density-wave
theory and the breakdown of the Kondo effect
\cite{pepin1,pepin}. Indeed, $S_T$ is sensitive to the
derivative of the density of electronic states and the change in
the relaxation time at $\mu$ \cite{lifshitzem:1980}. Using the
Boltzmann equation, the thermopower $S_T$ can be written as
\cite{beh,miy,zlat}
\begin{equation}
S_T=-\frac{\pi^2k_B^2T}{3e}\left[\frac{\partial\ln\sigma(\varepsilon)}
{\partial\varepsilon}\right]_{\varepsilon=\mu},\label{SB}
\end{equation}
where $k_B$ and $e$ are, respectively, the Boltzmann constant
and the elementary charge, while $\sigma$ is the dc electric
conductivity of the system, given by
\begin{equation}
\sigma(\varepsilon)=2e^2\tau(\varepsilon)\int\delta(\mu-\varepsilon({
p}))v({ p})v({ p})\frac{d{ p}}{(2\pi)^3},\label{SSG}
\end{equation}
where ${ p}$ is the electron wave-vector, $\tau$ is the
scattering time, and  $v$ denotes the velocities of electrons
belonging to the bands crossing the Fermi surface at
$\varepsilon=\mu$. Thus, we see from Eq. \eqref{SSG} that the
thermoelectric power $S_T$ is sensitive to the derivative of the
density of electronic states $N(\varepsilon=\mu)$ and the change
in the relaxation time at $\varepsilon=\mu$. On the basis of the
Fermi liquid theory description, the term in the brackets on the
right hand side of Eq. \eqref{SB} can be simplified, so that one
has $S_T\propto T N(\varepsilon=\mu)\propto C\propto M^*T$ at
$T\to0$ \cite{beh,miy,zlat}. As a result, under general
conditions and upon taking into account that charge and heat
currents at low temperatures are transported by quasiparticles,
the ratio $(S_T/C)\simeq const$ \cite{beh,miy,zlat}. It is seen
from Fig. \ref{figYb} that in the case of $\mathrm{YbRh_2Si_2}$
and at $T\geq T_{NL}$, the isotherms $-S_T(B)/T$ behave like
$C/T$: They exhibit a broad maximum that sharpens and shifts to
lower fields upon cooling \cite{TP,TPJ}. Here $T_{NL}$ is the
temperature of antiferromagnetic (AF) ordering ($T_{NL}=70$ mK)
at a critical field $B_{c0}=60$ mT, applied perpendicular to the
magnetically hard c axis \cite{shall}. Thus, $S_T/T\propto
C/T\propto\chi\propto M^*$ over a wide range of $T$, since in
the framework of FC theory, quasiparticles are responsible for
the thermodynamic and transport properties. It is worth noting
that $S_T/T\propto M^*$ in a low-disorder two-dimensional
electron system in silicon, and tends to a diverge at a finite
disorder-independent density \cite{shash}, thus confirming that
the charge and heat currents are transported by quasiparticles.
\begin{figure}[!ht]
\begin{center}
\includegraphics [width=0.47\textwidth]{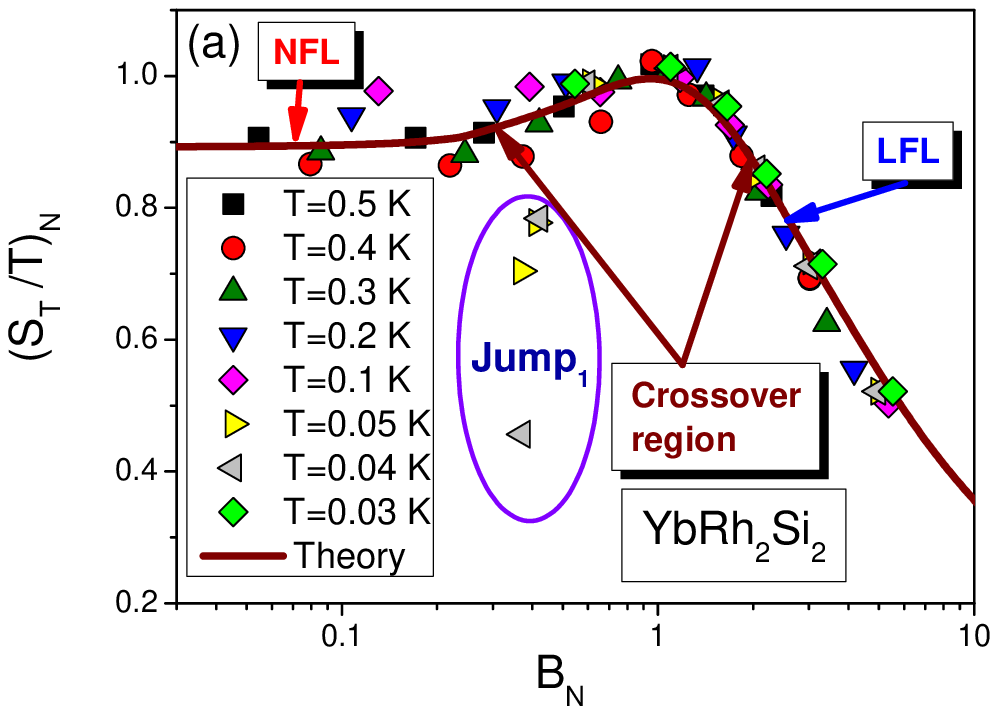}
\includegraphics [width=0.47\textwidth]{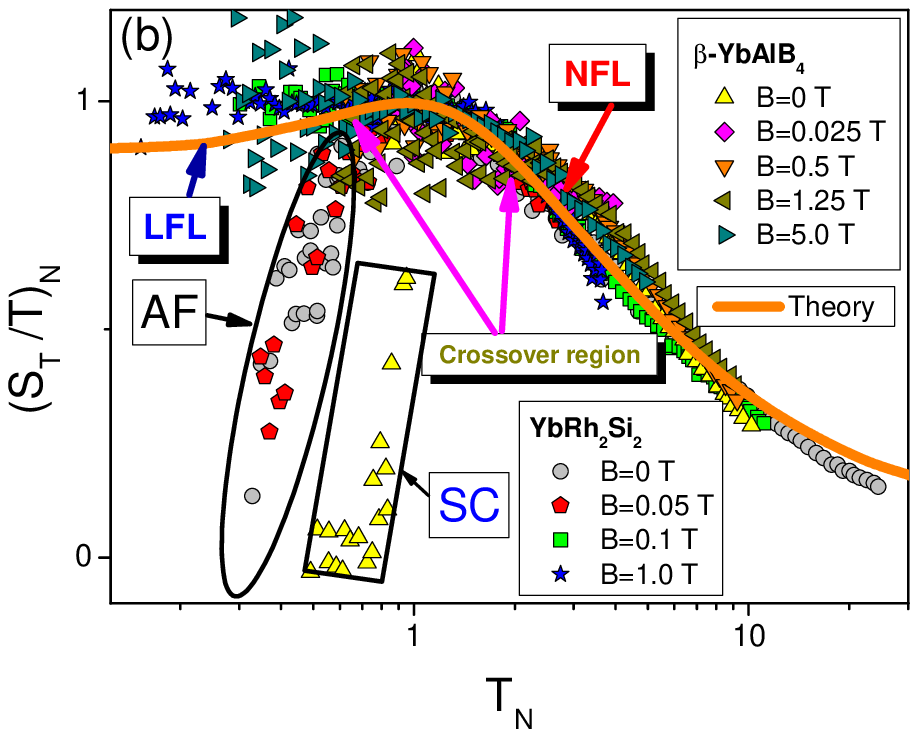}
\end{center}
\caption{(color online) (a) Normalized isotherm $(S_T(B)/T)_N$
versus normalized magnetic field $B_N$ for different
temperatures shown in the legend. The LFL behavior takes place
at $B_N>1$. (b) Temperature dependence of the normalized
thermopower $(S_T/T)_N$ under several magnetic fields shown in
the legend. The experimental data are extracted from
measurements on $\rm{YbRh_2Si_2}$ \cite{TP,TPJ} and on
$\beta$-${\rm YbAlB_4}$\, \cite{machida}. As it is explained in
the text, the data, taken at the AF phase \cite{TP,TPJ} and at
the superconducting one (SC) \cite{machida} and confined by both
the ellipse and the rectangle, respectively, violate the scaling
behavior. The solid curves in (a) and (b) represent calculated
$(C/T)_N$ displayed in Fig \ref{FIG2} (b).} \label{fig3}
\end{figure}

To reveal the universal scaling behavior of the thermopower
$S_T/T\propto C/T\propto M^*$, we normalize $S_T/T$ in the same
way as in the normalization of $C/T$: the normalized function
$(S_T/T)_N$ is obtained by normalizing $(S_T/T)$ by its maximum
value, occurring at $T=T_M$, and the temperature $T$ is scaled
by $T_M$. Taking into account that $S_T/T\propto C/T$\,
\cite{beh,miy,zlat}, we conclude that $(S_T/T)_N=(C/T)_N=M^*_N$,
provided that the system in question is located away from
possible phase transitions. This universal function
$(C/T)_N=M^*_N$ is displayed in Fig.~\ref{FIG2} (b). Figures
\ref{fig3} (a) and (b) report $(S_T/T)_N$ as a function of the
normalized magnetic field $B_N$ and $T_N$, respectively. In Fig.
\ref{fig3} (a), the function $(S_T/T)_N$ is obtained by
normalizing $(S_T/T)$ by its maximum occurring at $B_M$, and the
field $B$ is scaled by $B_M$. As seen from Eq. \eqref{UN2}, the
LFL behavior takes place at $B_N>1$, since $(S_T/T)_N=M^*_N$,
and $M^*_N\propto (B-B_{c0})^{-2/3}$ are $T$-independent, while
at $B_N<1$, $M^*_M$ becomes $T$-dependent and exhibits the NFL
behavior with $M^*_N\propto T^{-2/3}_N$. It is seen from Figs.
\ref{fig3} (a) and (b) that the calculated values of the
universal function $M^*_N$ are in good agreement with the
corresponding experimental data \cite{TP,TPJ,machida} over the
wide range of the normalized magnetic field. Thus, $(S_T/T)_N$
exhibits the universal scaling behavior over a wide range of its
scaled variable $B_N$ and $T_N$. Figure \ref{fig3} (a) also
depicts a violation of the scaling behavior for $B\leq B_{c0}$
when the system enters the AF phase. Moreover, as seen from
Figs. \ref{figYb} and \ref{fig3} (a) and (b), the scaling
behavior is violated at $T\leq T_{NL}$ by two downward jumps. It
is shown, these two jumps reflect the presence of flat band at
$\mu$ in the single particle spectrum $\varepsilon({ p})$ of
heavy electrons in $\rm{YbRh_2Si_2}$ \cite{shag2016}. In the
same way, as it is seen from Fig. \ref{fig3} (b), the scaling
behavior is violated by the superconducting (SC) phase
transition, taking place in $\beta$-${\rm YbAlB_4}$ at $T_c=$ 80
mK \, \cite{machida}.

\begin{figure}[!ht]
\begin{center}
\includegraphics [width=0.47\textwidth]{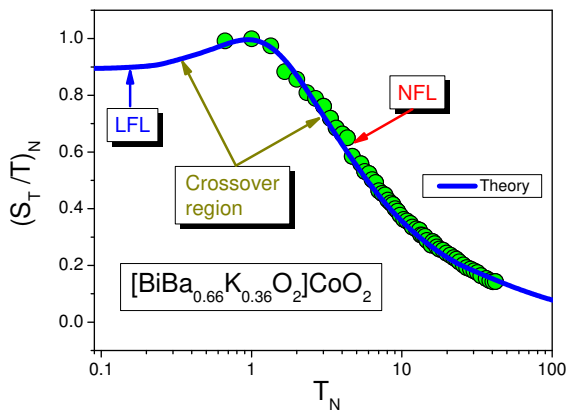}
\end{center}
\caption{(color online) Temperature dependence of $(S_T(T)/T)_N$
at magnetic field $B=0$, extracted from measurements on $\rm
[BiBa_{0.66}K_{0.36}O_{2}]CoO_{2}$\, \cite{limel}, is displayed
versus $T_N$. The solid line is the same as one depicted in Fig.
\ref{FIG2} (b).} \label{fig6}
\end{figure}
We now show that the observed scaling behavior of $(S_T/T)_N$ is
universal by analyzing experimental data on the thermopower for
$\rm [BiBa_{0.66}K_{0.36}O_{2}]CoO_{2}$ \cite{limel}. By
plotting $(S_T/T)_N$ as a function of $T_N$ in Fig. \ref{fig6},
the universal scaling behavior and the three regimes are seen to
be in a complete agreement with the reported overall behavior in
both Figs. \ref{FIG2} (b) and \ref{metaSrRuO} (a), and Figs.
\ref{fig3} (a), (b) as well.

Thus, we have revealed and explained the universal scaling
behavior of the thermopower $S_T/T$ in such different HF
compounds as $\rm{YbRh_2Si_2}$, $\beta$-${\rm YbAlB_4}$, and
$\rm [BiBa_{0.66}K_{0.36}O_{2}]CoO_{2}$. Our calculations are in
good agreement with experimental observations, and demonstrate
that $S_T/T$ exhibits the universal scaling behavior that
characterizes and singles out the new state of matter.

\section{Two-dimensional $\rm ^3He$}\label{3He}

In Subsection \ref{hf}, it was discussed that the electronic
system of HF metals demonstrates the universal low-temperature
behavior irrespectively of their magnetic ground state.
Therefore it is of crucial importance to check whether this
behavior can be observed in 2D Fermi systems. Fortunately, the
measurements on 2D $\rm ^3He$ are available
\cite{casey:1998,neumann:2007}. These measurements are extremely
significant as they allow one to check for the universal
behavior in the system formed by $\rm ^3He$ atoms, which are
essentially different from electrons \cite{shaginyan:2008}.
Namely, atoms of 2D $\rm ^3He$ are neutral fermions with spin
$S=1/2$. They interact with each other by van-der-Waals forces
with strong hardcore repulsion and a weakly attractive tail. The
different character of interparticle interaction along with the
fact, that the mass of $\rm ^3He$ atom is three orders of
magnitude larger than that of an electron, makes $\rm ^3He$ to
have drastically different microscopic properties
\begin{figure} [! ht]
\begin{center}\vspace*{-0.5cm}
\includegraphics [width=0.45\textwidth]{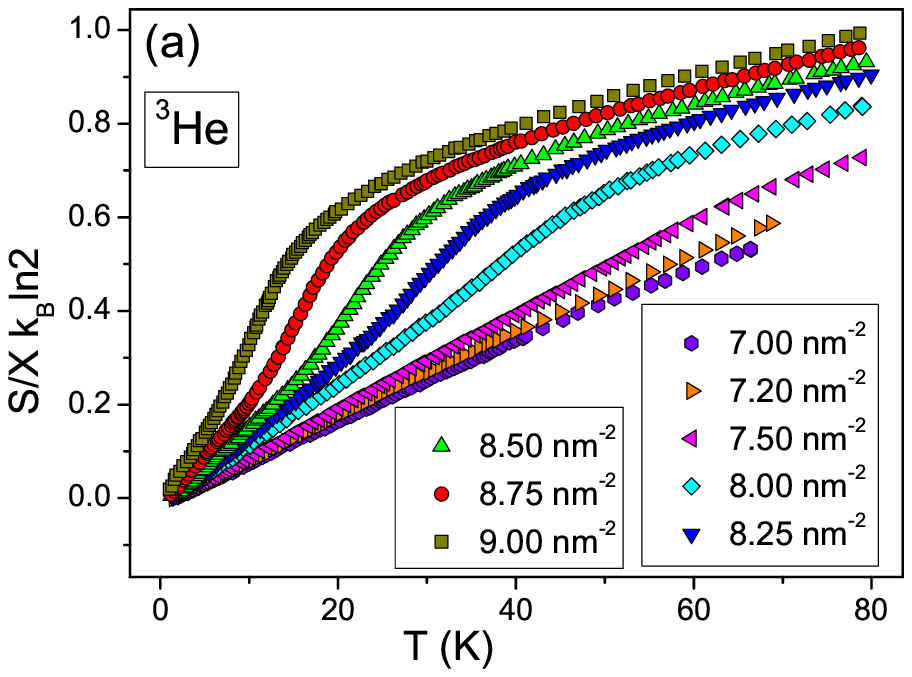}
\includegraphics [width=0.47\textwidth]{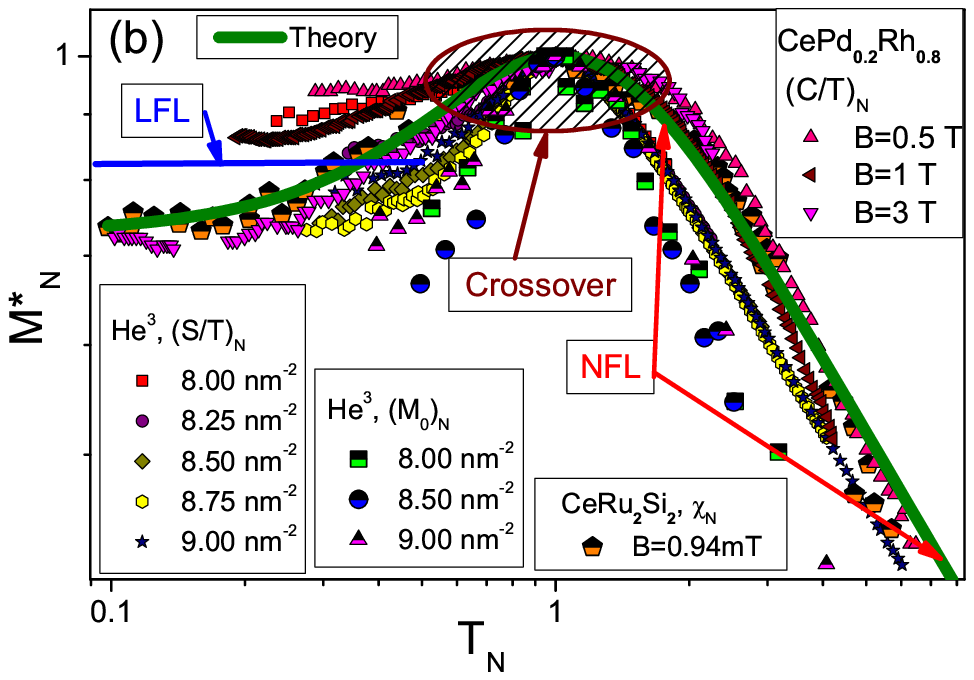}
\end{center}\vspace*{-0.5cm}
\caption{(color online) Universal scaling of $\rm^3He$. Panel
(a): Temperature dependence of the entropy $S$ of 2D $\rm^3He$
at different densities $x$ (shown in the legends) versus $T$
\cite{neumann:2007}. Panel (b): The normalized effective mass
$M^*_N$ as a function of the normalized temperature $T/T_{\rm
max}$ is extracted from experimental data using Eq.
\eqref{NORM}. The behavior of $M^*_N$ is extracted from
experimental data for $S(T)/T$ in 2D $^3$He \cite{neumann:2007}
and 3D HF compounds with different magnetic ground states, such
as $\rm{CeRu_2Si_2}$ and $\rm CePd_{1-x}Rh_x$
\cite{oeschler:2008,takahashi:2003}, and fitted by the universal
function (\ref{UN2}).}\label{3he2d}
\end{figure}
than that of three-dimensional (3D) HF metals that constitute
the most essential for our consideration. Because of this
drastic difference one can not be sure that the macroscopic
properties of the two fermion systems, mentioned above, will be
similar. The bulk 3D liquid $\rm ^3He$ is historically the first
object, to which the LFL theory has been applied. This
substance, being intrinsically isotropic Fermi-liquid with
negligible spin-orbit interaction is an ideal object to test the
LFL theory. { Unfortunately, FCQPT cannot be observed in 3D
liquid $\rm ^3He$, for it undergoes the first order phase
transition under the application of pressure $P$ long before
some deviations from the LFL theory is observed. In contrast to
3D $\rm ^3He$, its 2D counterpart exhibits strong deviations
from the LFL theory as soon as its number density $x$ approaches
the critical density $x_c$, as it is seen from Fig. \ref{3he2d},
panel (a). Thus, we can explore the behavior of 2D $\rm ^3He$
that is not hidden by possible phase transitions at $x<x_c$.}

2D films of $\rm ^3He$ have been fabricated and thermodynamic
properties have been thoroughly investigated
\cite{casey:1998,neumann:2007}. Among these properties are
measurements of $M^*$ as a function of the number density $x$
and the entropy $S$ as a function of $x$ versus $T$, displayed
in Figs. \ref{Fig6} and \ref{3he2d}, panel (a), respectively. We
conclude that 2D $\rm ^3He$ is near FCQPT, since $M^*$ diverges
at $x\to x_c$, and at $x\gtrsim 7.50$ nm$^{-2}$ the entropy is
no more a linear function of temperature $T$, exhibiting the NFL
behavior. Now we use the universal behavior of $M^*_N$ given by
Eq. \eqref{UN2} to fit the experimental data collected on both
2D $^3$He and 3D HF metals \cite{amusia:2015}. $M^*_N$,
extracted from the measurements on the $^3$He film
\cite{neumann:2007} at different densities $x<x_c$ smaller than
at the critical point $x_c=9.9 \pm 0.1$ nm$^{-2}$ is reported in
Fig. \ref{3he2d}, panel (b). In this panel, $M^*_N$ extracted
from the specific heat of ferromagnet CePd$_{0.2}$Rh$_{0.8}$
\cite{oeschler:2008} and from AC magnetic susceptibility of
paramagnet CeRu$_2$Si$_2$ \cite{takahashi:2003} are plotted for
different magnetic fields. It is seen that the universal
behavior of $M^*_N$ of 2D $^3$He quasiparticles given by Eq.
\eqref{UN2} (solid curve in Fig. \ref{3he2d}, panel (b)) is in
agreement with experimental data. All the samples of 2D $^3$He
are located at $x<x_c$, where the system progressively disrupts
its LFL behavior at elevated temperatures. In that case the
control parameter, driving the system towards its critical point
$x_c$, is merely the number density $x$. It is seen that the
behavior of $M^*_N$, extracted from measurements on 2D $^3$He,
is almost the same that of the 3D HF compounds
\cite{shaginyan:2008}. We conclude that Eq. \eqref{UN2} allows
us to reduce a four variable function describing the effective
mass to a function of a single variable. Indeed, the effective
mass depends on magnetic field, temperature, number density, and
on the composition of the corresponding HF compound. It follows
from our consideration that all these variables can be merged in
the single variable by using Eq. \eqref{UN2}. We conclude that
despite absolutely different microscopic nature of 2D $^3$He and
3D HF metals, they exhibit the new state of matter,
demonstrating the universal scaling.

\section{Quantum spin liquid}\label{QSL}

The first experimental observation of quantum spin liquid (QSL),
supporting exotic spin excitations - spinons - and carrying
fractional quantum numbers in the herbertsmithite $\rm
ZnCu_3(OH)_6Cl_2$, is reported in Ref. \cite{han:2012}. QSL can
be viewed as an exotic quantum state composed of hypothetical
particles such as fermionic spinons which carry spin $1/2$ and
no charge. The herbertsmithite $\rm ZnCu_3(OH)_6Cl_2$ has been
identified as a $S=1/2$ Heisenberg antiferromagnet on a perfect
kagome lattice, see Ref. \cite{bert:2010} for a recent review.
In $\rm ZnCu_3(OH)_6Cl_2$, the $\rm Cu^{2+}$ ions with $S=1/2$
form the triangular kagome lattice, and are separated by
nonmagnetic intermediate layers of $\rm Zn$ and $\rm Cl$ atoms.
The planes of the $\rm Cu^{2+}$ ions can be considered as
two-dimensional (2D) layers with weak magnetic interactions
along the third dimension. A simple kagome lattice has a
dispersionless topologically protected branch of the spectrum
with zero excitation energy that is the flat band
\cite{green:2010,heikkila:2011}. In that case FCQPT forms a
strongly correlated quantum spin liquid (SCQSL) composed of
fermions with zero charge, $S=1/2$, and the effective mass
$M^*_{\rm mag}$, occupying the corresponding Fermi sphere with
the Fermi momentum $p_F$
\cite{shaginyan:2012,shaginyan:2011,han:2012:A,devries:2008}.
\begin{figure} [! ht]
\begin{center}
\vspace*{-0.2cm}
\includegraphics [width=0.35\textwidth]{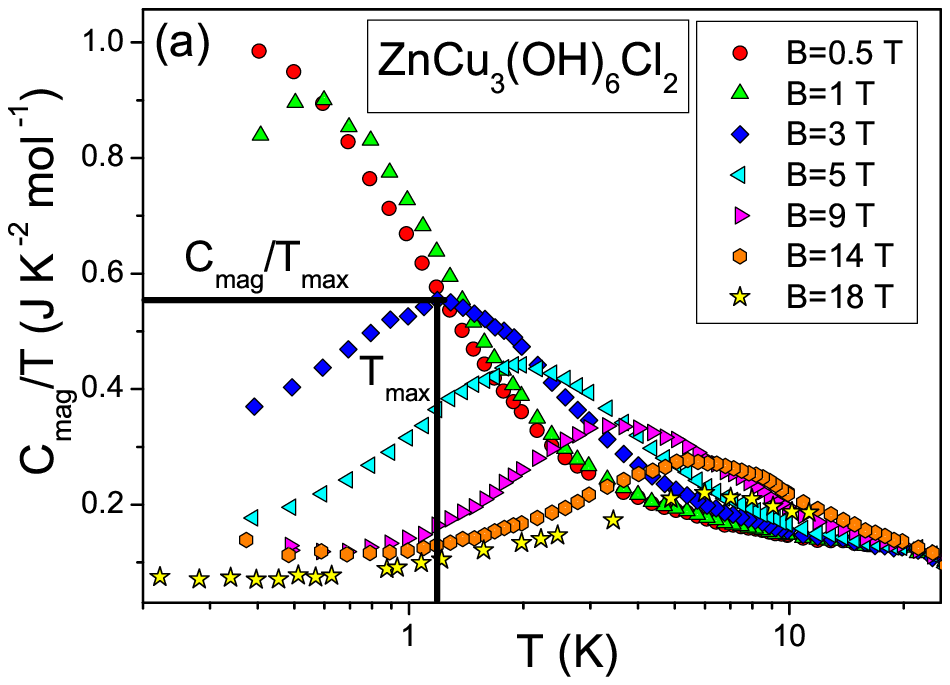}
\includegraphics [width=0.35\textwidth]{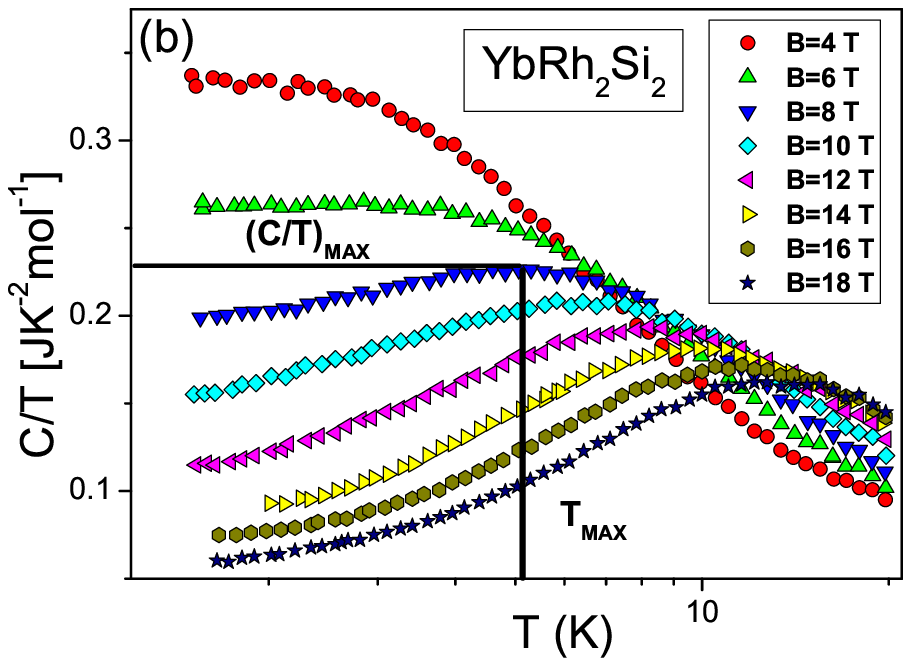}
\includegraphics [width=0.35\textwidth]{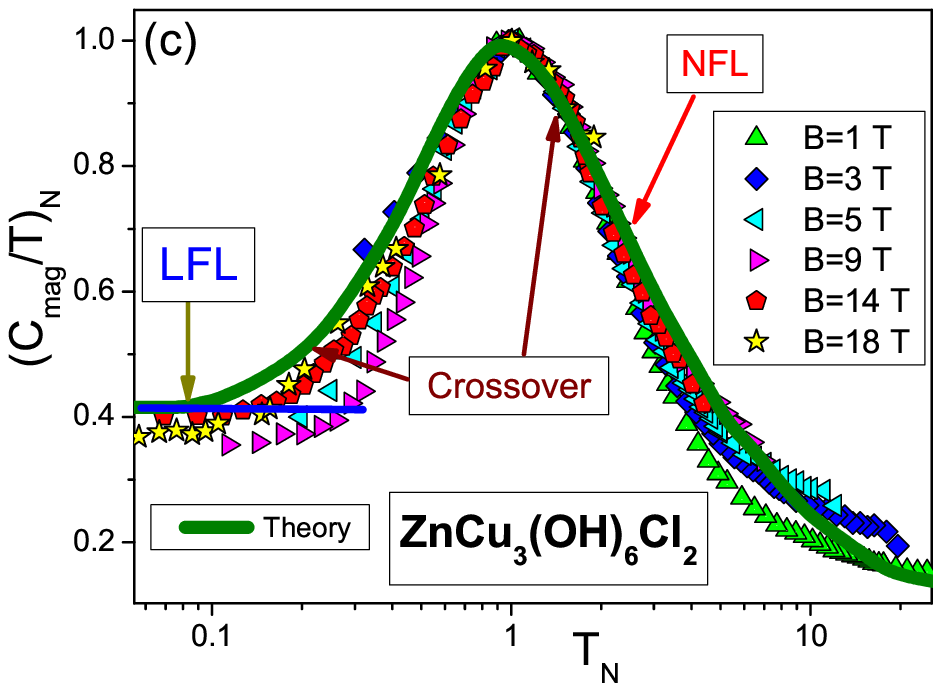}
\end{center}
\vspace*{-0.3cm} \caption{(color online) The specific heat
$C_{mag}/T$ of the $\rm ZnCu_3(OH)_6Cl_2$ insulator compared to
that of the archetypical HF metal $\rm YbRh_2Si_2$. Panel (a):
The specific heat $C_{\rm mag}/T$ of the QSL is extracted from
measurements of the specific heat on $\rm ZnCu_3(OH)_6Cl_2$ at
different magnetic fields shown in the legend
\cite{devries:2008,gegenwart:2006:A}. Panel (b) reports the
$T$-dependence of the electronic specific heat $C/T$ of $\rm
YbRh_2Si_2$ at different magnetic fields
\cite{devries:2008,gegenwart:2006:A} as shown in the legend.
Panel (c): Scaling behavior of the normalized specific heat
$(C_{\rm mag}/T)_N$ extracted from the measurements shown in the
panel (a). The LFL, crossover and NFL regions are shown by the
arrows. { The calculations are represented by the solid curve,
describing the normalized effective mass $M^*_N$ at high
magnetic fields extracted from the specific heat (C/T)
measurements of $\rm YbRh_2Si_2$ shown in the panel (b)
\cite{shaginyan:2011:C}.}} \label{FIG4}
\end{figure}

A comparison of the QSL specific heat $C_{\rm mag}/T\propto
M^*_{\rm mag}$, extracted from measurements on $\rm
ZnCu_3(OH)_6Cl_2$ \cite{han:2012:A} with $C/T$ of $\rm
YbRh_2Si_2$ \cite{devries:2008,gegenwart:2006:A,shaginyan:2012},
is shown in Fig. \ref{FIG4}, panels (a) and (b), respectively.
As seen from Fig. \ref{FIG4}, panel (c), the QSL exhibits the
scaling behavior similar to that of the electron liquid in HF
metals under the application of magnetic fields (see Figs.
\ref{FIG2}, panel (a), and \ref{metaSrRuO}, panels (a,b)) that
lead to full polarization of the corresponding subband
\cite{amusia:2015}. As a result, the specific heat becomes about
half of that in small magnetic fields, as it is seen from panel
(c) of Fig. \ref{FIG4} and Fig. \ref{FIG2}, panel (b)
\cite{amusia:2015}. From the same Figures one can see that the
LFL, crossover and NFL regions are presented in both $\rm
ZnCu_3(OH)_6Cl_2$ and $\rm YbRh_2Si_2$. The striking feature of
the specific heat behavior is the strong dependence on the
magnetic field seen from panels (a) and (b) of Fig. \ref{FIG4}.
{ Both $C_{\rm mag}/T$ of $\rm ZnCu_3(OH)_6Cl_2$ and $C/T$ of
$\rm YbRh_2Si_2$ exhibit the same qualitative and even
quantitative behavior that allows us to view the herbertsmithite
as insulator of new type with SCQSL, that occupies the Fermi
sphere of a finite volume, as that of $\rm YbRh_2Si_2$ occupied
by heavy-electron liquid
\cite{shaginyan:2011:C,shaginyan:2012,shaginyan:2011,shaginyan:2012:A}.
It is worthy to note that the thermodynamic properties of $\rm
YbRh_2Si_2$ in strong magnetic fields can be explained within
the frame of the FC theory  \cite{shaginyan:2011:C}, and these
turn out to be similar to that of $\rm ZnCu_3(OH)_6Cl_2$, as it
is seen from Fig. \ref{FIG4}, panels (a), (b), and (c). We
recall that at low temperatures the heat capacity of common
insulators $C\propto T^3$, and $C/T\to 0$ that is in vivid
contrast with the above experimental and theoretical
observation.}

\begin{figure} [! ht]
\begin{center}
\includegraphics [width=0.33\textwidth]{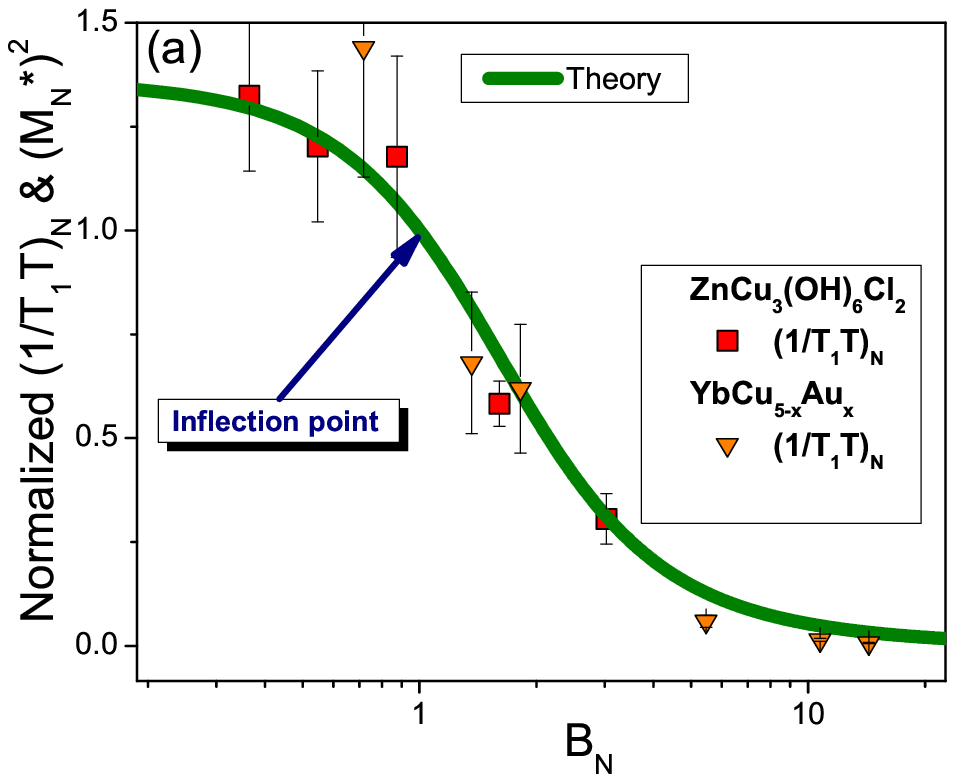}
\includegraphics [width=0.33\textwidth]{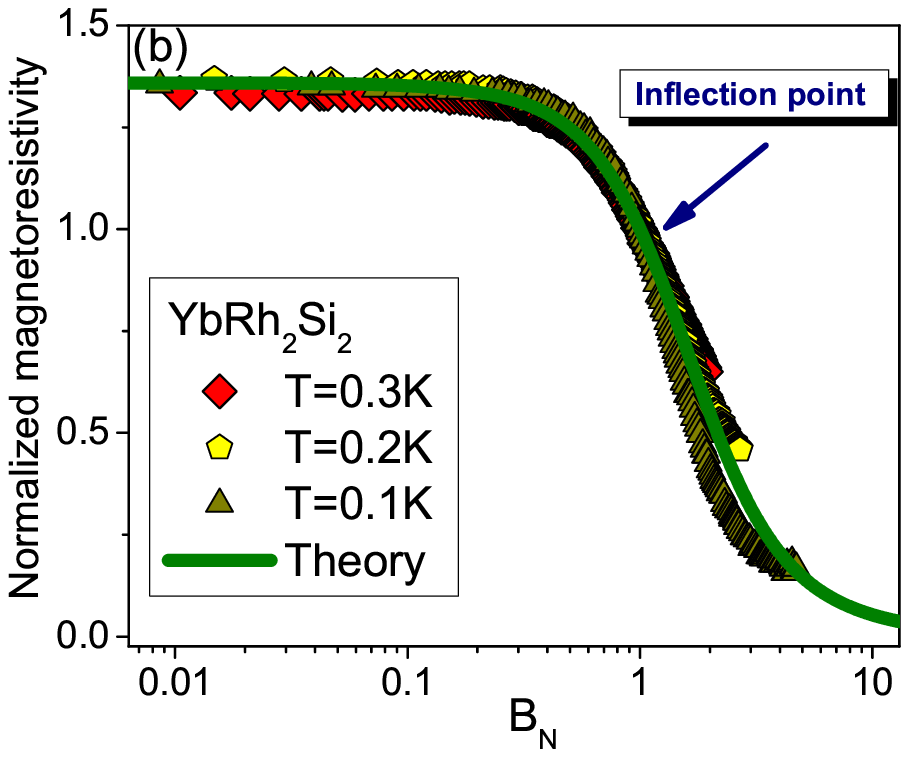}
\includegraphics [width=0.33\textwidth]{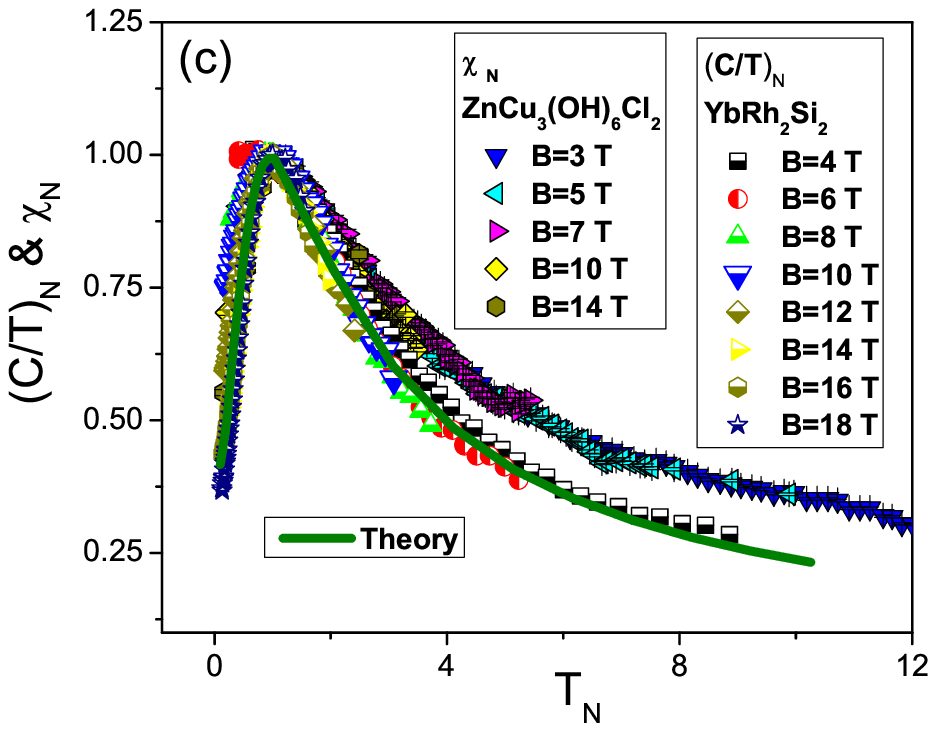}
\vspace*{-0.5cm}
\end{center}
\vspace*{-0.3cm} \caption{(color online) Relaxation, transport
and thermodynamic properties of QSL. Panel (a): comparison of
the relaxation properties of the herbertsmithite with those of a
HF metals. The normalized spin-lattice relaxation rate
$(1/T_1T)_N$ at fixed temperature as a function of magnetic
field: Solid squares correspond to data on $(1/T_1T)_N$
extracted from measurements on $\rm ZnCu_3(OH)_6Cl_2$
\cite{imai:2008}, while the solid triangles correspond to those
extracted from measurements on $\rm YbCu_{5-x}Au_{x}$ with
$x=0.4$ \cite{carretta:2009}. The inflection point, representing
the crossover region where the normalization is taken, is shown
by the arrow. Our calculations based on Eqs. \eqref{LF1} and
\eqref{UN2} are depicted by the solid curve, tracing the scaling
behavior of $(M^*_N)^2$. Panel (b): The normalized longitudinal
magnetoresistance $\rho_N$ versus $B_N$. $\rho_N$ is extracted
from measurements on $\rm YbRh_2Si_2$ at different temperatures
\cite{gegenwart:2007} that are listed in the legend. The solid
curve represents our calculations and coincides with that shown
in the panel (a). Panel (c): The normalized effective mass
$M^*_N$ extracted from measurements of susceptibility $\chi$
\cite{helton:2010} on $\rm ZnCu_3(OH)_6Cl_2$ and $(C/T)$ on $\rm
YbRh_2Si_2$ \cite{gegenwart:2006:A}. Our calculations of $M^*_N$
are shown by the solid curve.}\label{T1}
\end{figure}
Fig, \ref{T1} (a) displays the normalized spin-lattice
relaxation rates $(1/T_1T)_N$ at fixed temperature versus
normalized magnetic field $B_N$. It is seen from Fig. \ref{T1},
that the magnetic field $B$ progressively reduces $1/T_1T$, and
the spin-lattice relaxation rate as a function of $B$ possesses
an inflection point at some $B=B_{inf}$ shown by the arrow. To
clarify the universal scaling behavior of QSL in the
herbertsmithite  and in HF metals, we normalize both the
function $1/T_1T$ and the magnetic field. Namely, we normalize
$(1/T_1T)$ by its value at the inflection point, and the
magnetic field is normalized by $B_{inf}$, $B_N=B/B_{inf}$.
Since $(1/T_1T)_N=(M^*_N)^2$
\cite{shaginyan:2010,shaginyan:2009:D}, we expect that different
strongly correlated Fermi systems located near FCQPT will
exhibit the same behavior of the normalized spin-lattice
relaxation rate. It is seen from Fig. \ref{T1}, that both the
herbertsmithite $\rm ZnCu_3(OH)_6Cl_2$ \cite{imai:2008} and HF
metal $\rm YbCu_{5-x}Au_{x}$ \cite{carretta:2009} demonstrate
similar behavior of the normalized spin-lattice relaxation rate.
As seen from Fig. \ref{T1}, panel (a), at $B\leq B_{inf}$ (or
$B_N\leq1$), when the system is in its NFL region, the
normalized relaxation rate $(1/T_1T)_N$ depends weakly on the
magnetic field, while at higher fields, as soon as the system
enters the LFL region, $(1/T_1T)_N=(M_N^*)^2\propto B^{-4/3}$
diminishes in agreement with Eq. \eqref{UN2}; in that case
$n=8/3$ \cite{shaginyan:2010,amusia:2015}. Figure \ref{T1},
panel (b) displays the normalized longitudinal magnetoresistance
(LMR) $\rho_N$ at fixed temperatures versus the normalized
magnetic field $B_N$ taken on the HF metal $\rm YbRh_2Si_2$
\cite{gegenwart:2007}. Fig. \ref{T1}, panel (b), shows that the
growing magnetic field progressively reduces LMR, and it as a
function of $B$ it possesses an inflection point at $B=B_{inf}$
shown by the arrow. Both the normalized LMR and $(1/T_1T)_N$
obey the same equation \cite{shaginyan:2010,amusia:2015}
\begin{equation}\label{43}
\frac{1}{(T_1T)_N}=\rho_N(B_N)=\frac{\rho_{LMR}(B_N)-\rho_0}
{\rho_{inf}}=(M^*_N)^2\propto B^{-4/3},
\end{equation}
where $\rho_0$ is the residual resistance, $\rho_{inf}$ is LMR
taken at the inflection point, $\rho_{LMR}$ is LMR, $\rho_N$ is
the normalized LMR, and $B_N=B/B_{inf}$. We normalize LMR by
their values at the inflection point, and the magnetic field is
normalized by $B_{inf}$, as it has been done in the case of the
spin-lattice relaxation rate. Thus, from Eq. \eqref{43} and Fig.
\ref{T1}, panels (a) and (b), we conclude that the application
of magnetic field $B$ leads to the crossover from the NFL to LFL
behavior and to the significant reduction in the relaxation rate
and LMR, while both the normalized LMR and the normalized
relaxation rate coincide, thus demonstrating the universal
scaling behavior given by Eq. \eqref{UN2}. From Fig. \ref{T1},
panel (c) it is clearly seen that the data collected on both
$\rm ZnCu_3(OH)_6Cl_2$ \cite{helton:2010} and $\rm YbRh_2Si_2$
\cite{gegenwart:2006:A} merge into the same curve, respecting
scaling. This demonstrates that QSL of the herbertsmithite
behaves like HF liquid of $\rm YbRh_2Si_2$ in magnetic fields.
Thus, SCQSL of $\rm ZnCu_3(OH)_6Cl_2$ behaves like that of HF
metals, and signals that the herbertsmithite demonstrates the
new state of matter.

As  mentioned above, QSL plays a role of HF liquid embedded into
the insulating compound. Thus, we expect that the QSL in the
herbertsmithite to behave like the electronic liquid in HF
metals if the charge of an electron were zero. In that case, the
thermal resistivity $w$ of QSL is related to the thermal
conductivity $\kappa$ \cite{shaginyan:2013:D}
\begin{equation}\label{kap}
w=\frac{L_0T}{\kappa}=w_0+A_wT^2.
\end{equation}
The thermal resistivity $w$ behaves like the electrical
resistivity $\rho=\rho_0+A_{\rho}T^2$ of the electronic liquid,
since $A_w$ represents the contribution of spinon-spinon
scattering to the thermal transport, being analogous to the
contribution $A_{\rho}$ to the charge transport induced by
electron-electron scattering. Here, $L_0$ is the Lorenz number,
$\rho_0$ and $w_0$ are residual resistivity of electronic liquid
and QSL, respectively, and the coefficients are $A_w\propto
(M^*_{\rm mag})^2$ and $A_{\rho}\propto (M^*)^2$. Thus, we
expect that in the LFL region $A_w$ of the thermal resistivity
of QSL under the application of magnetic fields at fixed $T$ to
behave like $(1/T_1T)_N$ and $\rho_N$ shown respectively in Fig.
\ref{T1}, panels (a) and (b), namely $A_w\propto
(1/T_1T)_N\propto\rho_N\propto(M^*(B)_{\rm mag})^2$. However, in
the LFL region at fixed magnetic fields the thermal conductivity
$\kappa$ is a linear function of temperature, $\kappa\propto T$
\cite{shaginyan:2013:D}.

\begin{figure} [! ht]
\begin{center}
\vspace*{-0.5cm}
\includegraphics [width=0.47\textwidth]{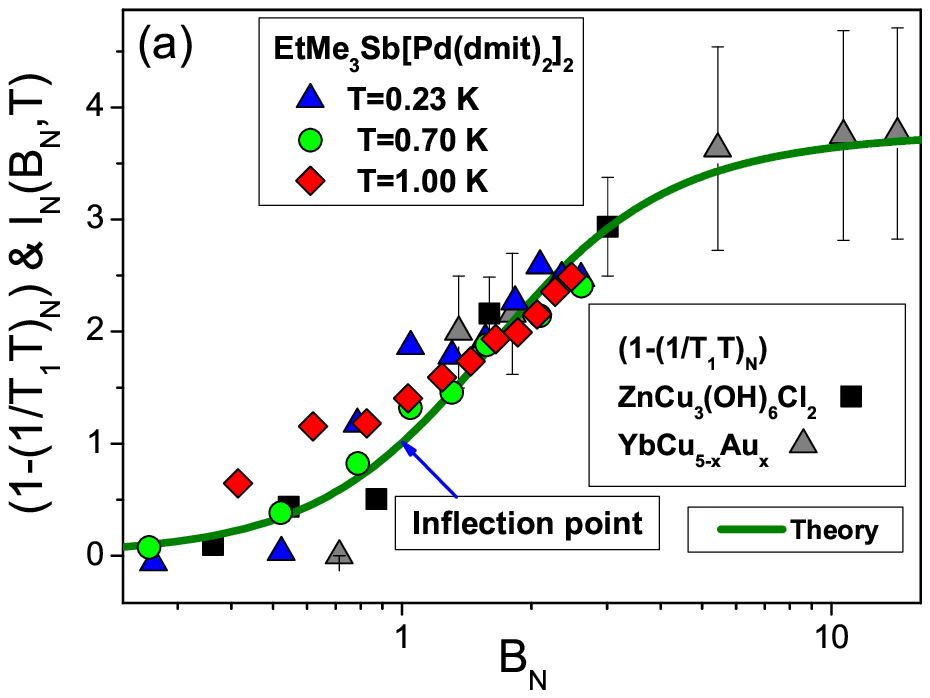}
\includegraphics [width=0.47\textwidth]{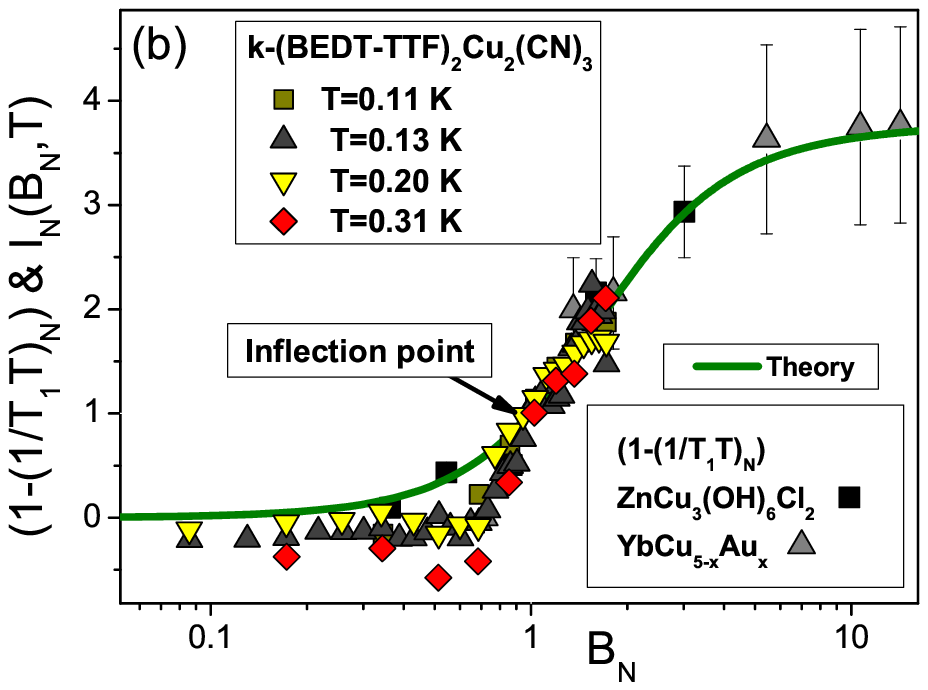}
\end{center}
\vspace*{-0.5cm} \caption{(color online) Normalized thermal
conductivity $I_N(B_N,T)$ of the organic insulators. Panel (a):
$I_N(B_N,T)$ measured on $\rm EtMe{_3}Sb[Pd(dmit)_{2}]_{2}$ as
function of the normalized magnetic field $B_N$ at temperatures
shown in the legend \cite{yamashita:2010,yamashita:2012}. The
inflection point is shown by the arrow. Our calculations based
on Eq. \eqref{TR} are represented by the solid curve. Panel (b):
$I_N(B_N,T)$ measured on $\rm\kappa-(BEDT-TTF)_2Cu_2(CN)_3$ as
function of the normalized magnetic field $B_N$ at temperatures
shown in the legend \cite{yamashita:2010,yamashita:2012}. The
inflection point is shown by the arrow. In both panels, the data
$(1-(1/T_1T)_N)$ extracted from measurements of the spin
relaxation rate on $\rm ZnCu_3(OH)_6Cl_2$ and $\rm YbCu_{5-x}Au_{x}$, 
see Fig. \ref{T1}, panel (a), are shown by
the filled squares and triangles. Our calculations based on Eq.
\eqref{TR} are indicated by the solid curve that is obtained
from the theoretical curve  shown in Fig. \ref{T1}, panels (a)
and (b) \cite{shaginyan:2013:D}.}\label{kappa1}
\end{figure}
The study of the thermal resistivity $w$ given by Eq.
\eqref{kap} leads to revelation of spinons as itinerant
excitations. The temperature dependence of thermal resistivity
$w$ represented by the finite term $w_0$ demonstrates that the
behavior of QSL is similar to that of metals, and there is a
finite residual term $\kappa/T$ in the zero-temperature limit of
$\kappa$. The presence of this term immediately establishes that
there are gapless excitations,  associated with the property of
normal metals, in which gapless electrons govern the heat
transport. The finite $w_0$ implies that in QSL both $k/T$ and
$C_{mag}/T\propto M^*_{\rm mag}$ remain nonzero as $T\to0$.
Therefore, gapless spinons, forming the Fermi surface, govern
the specific heat and the transport. Key information on the
nature of spinons is further provided by the $B$-dependence of
the coefficient $A_w$. The specific $B$-dependence of the
resistivity $w(B)\propto(M^*_{\rm mag})^2$, given by Eq.
\eqref{43}, would establish the behavior of QSL as SCQSL. We
note that the heat transport is polluted by the phonon
contribution. On the contrary, the phonon contribution is hardly
influenced by magnetic field $B$. Therefore, we expect the
$B$-dependence of the heat conductivity to be governed by
$A_w(B,T)$. Consider the approximate relation,
\begin{eqnarray}
\nonumber 1&-&\frac{A_w(B,T)}{A_w(0,T)}=
1-\left(\frac{M^*(B,T)_{\rm mag}}{M^*(0,T)_{\rm mag}}\right)^2\\
&\simeq&a(T)\frac{\kappa(B,T)-\kappa(0,T)}{\kappa(0,T)}\equiv
a(T)I(B,T),\label{TR}
\end{eqnarray}
where the coefficient $a(T)$ is $B$-independent. To derive
\eqref{TR}, we use Eq. \eqref{kap}, and obtain
\cite{amusia:2015,shaginyan:2013:D}
\begin{equation}\label{TR1}
\frac{\kappa}{L_0T}=\frac{1}{w_0+A_wT^2}+bT^2.
\end{equation}
Here, the term $bT^2$ describes the phonon contribution to the
heat transport. Upon carrying out simple algebra and assuming
that $[1-A_w(B,T)/A_w(0,T)]\ll 1$, we arrive at Eq. \eqref{TR}.
From Fig. \ref{T1}, panels (a) and (b), it is seen that the
effective mass $M^*_N(B)\propto M^*_{\rm mag}(B)$ is a
decreasing function of the magnetic field $B$. Then, it follows
from Eqs. \eqref{43} and \eqref{TR} that the function
$I(B,T)=[\kappa(B,T)-\kappa(0,T)]/\kappa(0,T)$ increases with
$B$ increasing.

Recent measurements of $\kappa(B)$ on the organic insulators
$\rm EtMe{_3}Sb[Pd(dmit)_{2}]_{2}$ and
$\rm\kappa-(BEDT-TTF)_2Cu_2(CN)_3$
\cite{yamashita:2010,yamashita:2012} are displayed in Fig.
\ref{kappa1}, panels (a) and (b). The measurements show that the
heat is carried by phonons and QSL, since the heat conductivity
is well fitted by $\kappa/T=a_1+a_2T^2$, where $a_1$ and $a_2$
are constants. The finite term $a_1$ implies that spinon
excitations are gapless in $\rm EtMe{_3}Sb[Pd(dmit)_{2}]_{2}$,
while in $\rm\kappa-BEDT-TTF)_2Cu_2(CN)_3$ gapless excitations
are still being debated \cite{yamashita:2012}. A simple
estimation indicates that the ballistic propagation of spinons
appears to be realized in the case of $\rm
EtMe{_3}Sb[Pd(dmit)_{2}]_{2}$
\cite{yamashita:2010,yamashita:2012}. From Fig. \ref{kappa1},
panels (a) and (b), the normalized data
$I(B,T)=[\kappa(B,T)-\kappa(B=0,T)]/\kappa(B=0,T)$ demonstrates
a strong $B$-dependence, namely the field dependence shows an
increase of thermal conductivity for increasing $B$. Such
behavior is in agreement with Eq. \eqref{43} and Fig. \ref{T1},
panel (a); it demonstrates that $(M^*(B)_{mag})^2$ is a
decreasing function of $B$. Consequently, it is seen from Eq.
\eqref{TR}, that $I(B,T)$ is an increasing function of $B$. It
is seen from Fig. \ref{kappa1}, panels (a) and (b), that
$I(B,T)$ as a function of $B$ possesses an inflection point at
some $B=B_{inf}$. To reveal the scaling behavior of QSL in both
$\rm EtMe{_3}Sb[Pd(dmit)_{2}]_{2}$ and
$\rm\kappa-(BEDT-TTF)_2Cu_2(CN)_3$, we normalize both the
function $I(B,T)$ and the magnetic field by their values at the
inflection points, as it was done in the case of $(1/T_1T)$, see
Fig. \ref{T1}, panel (a). In that case we eliminate factor
$a(T)$, appearing in Eq. \eqref{TR}, and our calculations are
free of any fitting parameters. Moreover, the theoretical curve
shown in Fig. \ref{kappa1}, panels (a) and (b), is extracted
from those shown in Fig. \ref{T1}, the panels (a) and (b), and,
thus, is defined by LMR measured on $\rm YbRh_2Si_2$ and by
$(1-(1/T_1T)_N)$ extracted from measurements of the spin
relaxation rate on $\rm ZnCu_3(OH)_6Cl_2$ and $\rm
YbCu_{5-x}Au_{x}$. Clearly, from Fig. \ref{kappa1}, the
normalized $I_N(B_N,T)$ exhibits the scaling behavior and
becomes a function of a single variable $B_N$, yielding good
overall agreement between our calculations and the experimental
data \cite{yamashita:2010,yamashita:2012}.

Neutron scattering measurements of the dynamic spin
susceptibility $\chi({ q},\omega,T)=\chi{'}({
q},\omega,T)+i\chi{''}({ q},\omega,T)$, as a function of
momentum $q$, frequency $\omega$ and temperature $T$, play
important role when identifying the properties of
quasiparticles.
\begin{figure} [! ht]
\begin{center}
\vspace*{-0.5cm}
\includegraphics [width=0.35\textwidth]{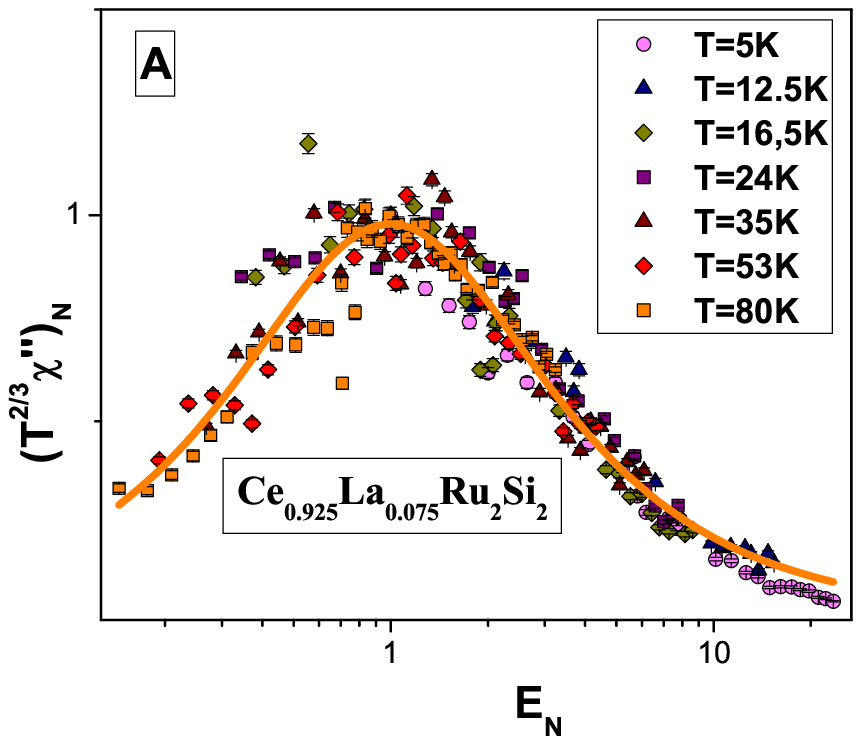}
\includegraphics [width=0.35\textwidth]{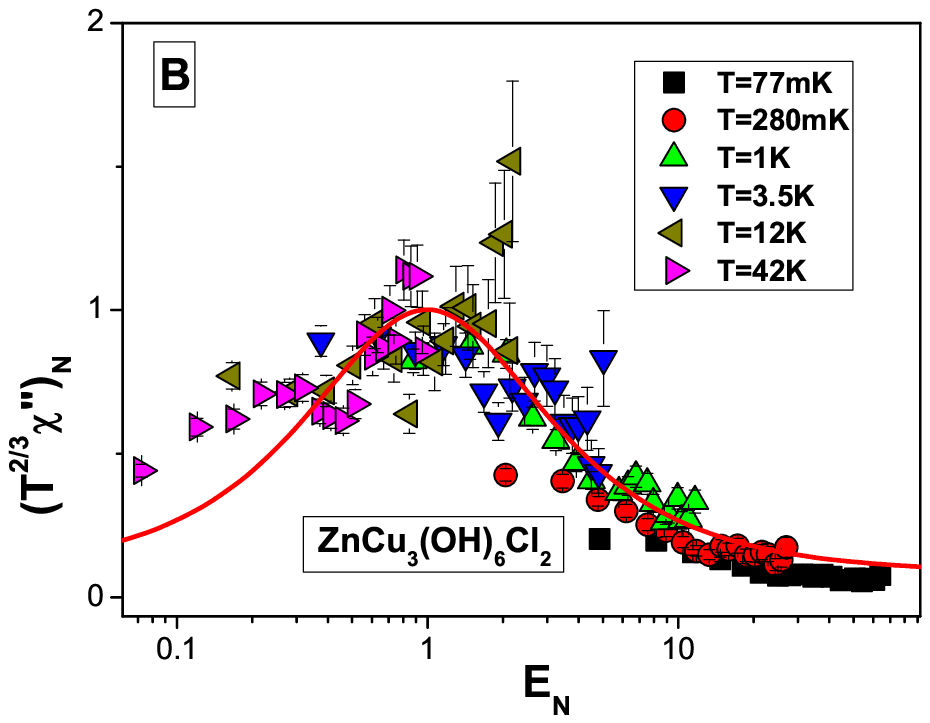}
\includegraphics [width=0.35\textwidth]{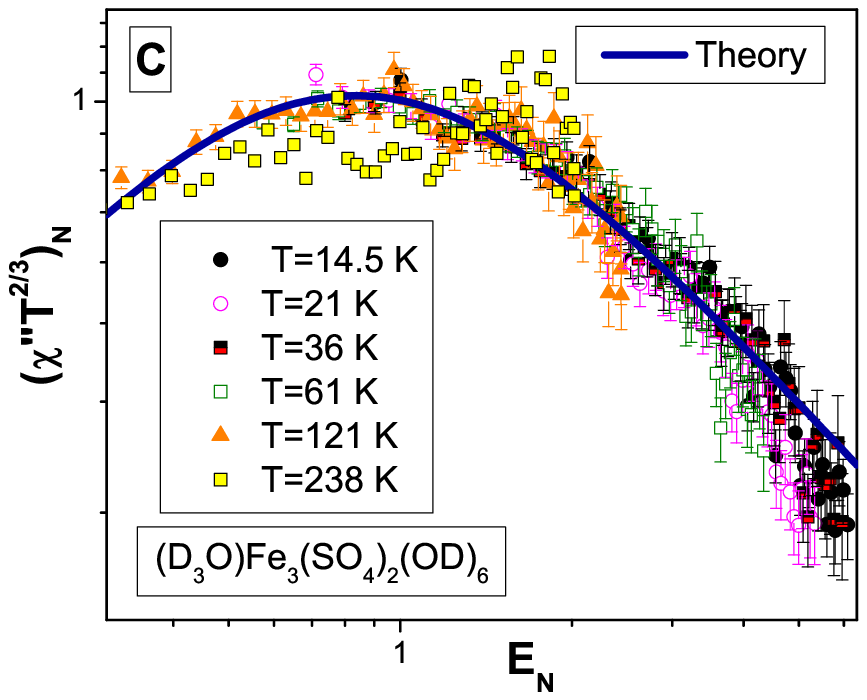}
\end{center}
\vspace*{-0.5cm} \caption{(color online) The scaling behavior of
the normalized dynamic spin susceptibility $(T^{2/3}\chi'')_N$.
Panel { A}: $(T^{2/3}\chi'')_N$ plotted against the
dimensionless variable $E_N$. The data are extracted from
measurements on the HF metal $\rm Ce_{0.925}La_{0.075}Ru_2Si_2$
\cite{knafo:2004}. Panel { B}: $(T^{2/3}\chi'')_N$ versus $E_N$.
The data are extracted from measurements on the herbertsmithite
$\rm ZnCu_3(OH)_6Cl_2$ \cite{helton:2010}. Panel { C}:
$(T^{2/3}\chi'')_N$ versus $E_N$. The data are extracted from
measurements on the deuteronium jarosite $\rm
(D_3O)Fe_3(SO_4)_2(OD)_6$ \cite{faak:2008}. For all these HF
compounds our calculations, based on Eq. \eqref{SCHIN}, are
shown by the solid curves \cite{shaginyan:2012:A}.}\label{FIG6}
\end{figure}
At low temperatures, such measurements reveal that
quasiparticles excitations of the new type insulators are
represented by spinons, form a continuum, and populate an
approximately flat band crossing the Fermi level
\cite{han:2012}. In that case it is expected that the normalized
susceptibility $(T^{2/3}\chi'')_{N}$ exhibits scaling as a
function of the dimensionless variable $E_N$
\cite{amusia:2015,shaginyan:2012:A}. As was done for $M^*_N$
when constructing Eq. \eqref{UN2}, we introduce the
dimensionless function
$(T^{2/3}\chi'')_{N}=T^{2/3}\chi''/(T^{2/3}\chi'')_{\rm max}$
and the dimensionless variable $E_N=E/E_{\rm max}$; then the
equation describing the normalized susceptibility
$(T^{2/3}\chi'')_{N}$ reads
\begin{equation}\label{SCHIN}
(T^{2/3}\chi'')_N\simeq\frac{b_1E_N}{1+b_2E_N^2},
\end{equation}
with $b_1$ and $b_2$ being fitting parameters which are adjusted
so that the function $(T^{2/3}\chi'')_{N}$ reaches its maximum
value 1 at $E_n=1$ \cite{amusia:2015,shaginyan:2012:A}. The
panel { A} of Fig. \ref{FIG6} reports $(T^{2/3}\chi'')_{N}$
extracted from measurements of the inelastic neutron scattering
spectrum on the HF metal $\rm Ce_{0.925}La_{0.075}Ru_2Si_2$
\cite{knafo:2004}. The scaled data obtained in measurements on
such a quite distinct strongly correlated systems as $\rm
ZnCu_3(OH)_6Cl_2$ \cite{helton:2010} and the deuteronium
jarosite $\rm (D_3O)Fe_3(SO_4)_2(OD)_6$ \cite{faak:2008} are
shown in the panel { B} and { C}. It is seen that our
calculations shown by the solid curves are in good agreement
with the experimental data collected on $\rm
Ce_{0.925}La_{0.075}Ru_2Si_2$, $\rm ZnCu_3(OH)_6Cl_2$ and $\rm
(D_3O)Fe_3(SO_4)_2(OD)_6$ over almost three orders of the scaled
variable $E_N$ and  $(T^{2/3}\chi'')_{N}$ does exhibit the
scaling behavior. Thus, the spin excitations in both $\rm
ZnCu_3(OH)_6Cl_2$ and $\rm (D_3O)Fe_3(SO_4)_2(OD)_6$ demonstrate
the same itinerate behavior as electron excitations of the HF
metal $\rm Ce_{0.925}La_{0.075}Ru_2Si_2$ do, and, therefore form
a continuum. This observation of the continuum is of great
importance since it clearly reveals the presence of QSL in the
herbertsmithite \cite{shaginyan:2012:A,shaginyan:2013:C}; this
was later confirmed by direct experimental observation
\cite{han:2012}.

Thus, the non-Fermi liquid behavior and the scaling properties
of such strongly correlated Fermi systems as insulators $\rm
ZnCu_3(OH)_6Cl_2$, $\rm (D_3O)Fe_3(SO_4)_2(OD)_6$, $\rm
EtMe{_3}Sb[Pd(dmit)_{2}]_{2}$,
$\rm\kappa-(BEDT-TTF)_2Cu_2(CN)_3$, and HF metals $\rm
Ce_{0.925}La_{0.075}Ru_2Si_2$, $\rm YbCu_{5-x}Au_{x}$, and $\rm
YbRh_2Si_2$ are described within the framework of the theory of
FC, and their scaled thermodynamic, transport and relaxation
properties are governed by Eqs. \eqref{UN2} and \eqref{SCHIN}.
Our calculations are in a good agreement with the experimental
data, and allow us to identify the low-temperature behavior of
$\rm ZnCu_3(OH)_6Cl_2$, $\rm (D_3O)Fe_3(SO_4)_2(OD)_6$,
$\rm\kappa-(BEDT-TTF)_2Cu_2(CN)_3$, and $\rm
EtMe{_3}Sb[Pd(dmit)_{2}]_{2}$ as determined by SCQSL. The same
behavior is observed in other HF compounds like HF metals and 2D
$^3$He. Thus, HF compounds with QSL can be viewed as a new type
of strongly correlated electrical insulators that form the new
state of matter and possess the properties and the universal
scaling behavior of HF metals with one exception: it resists the
flow of electric charge, but supports the spin current formed by
spinons.

\section{Quasicrystals}\label{QC}

The study of new materials dubbed quasicrystals (QCs) and
characterized by noncrystallographic rotational symmetry and
quasiperiodic translational properties have continued from quite
some time  \cite{shechtman:1984}. Study of such extraordinary HF
compounds as quasicrystals may shed light on the most basic
features related to the universal behavior observed in HF
compounds. Experimental measurements on the
gold-aluminium-ytterbium quasicrystal $\rm
Au_{51}Al_{34}Yb_{15}$ have revealed a specific behavior with
the unusual exponent $\beta\simeq0.51$ defining the divergency
of the magnetic susceptibility $\chi\propto T^{-\beta}$ at
$T\to0$ \cite{deguchi:2012}. The measurements have also revealed
that the observed NFL behavior transforms into that of the LFL
with application of a tiny magnetic field $B$. All these facts
challenge theory to explain the unique properties of the
gold-aluminum-ytterbium QC. In case of QCs electrons occupy a
new class of states denoted as "critical states", neither being
extended nor localized. Associated with these critical states,
characterized by an extremely degenerate confined wave function,
are the so-called "spiky" DOS \cite{fujiwara:1999}. These
predicted DOS are corroborated by experiments revealing the
spiky DOS \cite{widmer:2009}, which are formed by flat bands
\cite{trambly:2009}. As a result, we conclude that the
electronic system of some quasicrystals is located at FCQPT
without tuning \cite{amusia:2015,shaginyan:2013:A}.

\begin{figure} [! ht]
\begin{center}
\vspace*{-0.5cm}
\includegraphics [width=0.40\textwidth]{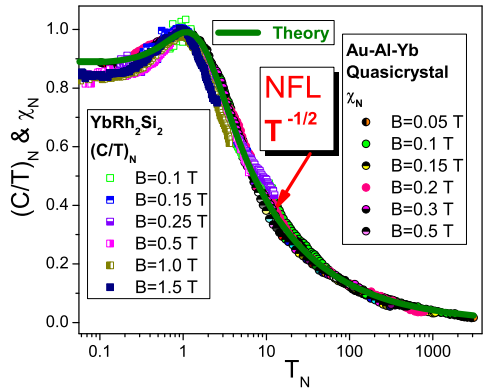}
\end{center}
\vspace*{-0.5cm} \caption{(color online) The normalized specific
heat $(C/T)_N$ and the normalized magnetic susceptibility
$\chi_N$ extracted from measurements in magnetic fields $B$ on
the HF metal $\rm YbRh_2Si_2$ \cite{oeschler:2008} and on the
quasicrystal $\rm Au_{51}Al_{34}Yb_{15}$ \cite{deguchi:2012},
respectively. The corresponding magnetic fields are listed in
the legends. The arrow shows the $T^{-1/2}$ regime taking place
at $T_N>1$. Our calculations are represented by the solid curve
tracing the scaling behavior of $(C/T)_N=\chi_N=M^*_N$ given by
Eq. \eqref{UN2}.}\label{FIG7_1}
\end{figure}
We now investigate the behavior of $\chi$ as a function of
temperature at fixed magnetic fields. The effective mass
$M^*(T,B)$ can be extracted from the data since $M^*(T,B)\propto
\chi$, see Eq. \eqref{NORM}. If the corresponding measurements
are carried out at fixed magnetic field $B$, then, as it follows
from the Eq. \eqref{UN2}, $\chi$ reaches the maximum $\chi_{M}$
at some temperature $T_{M}$. Upon normalizing both $\chi$ and
the specific heat $C/T$ by their peak values at each value of
magnetic field $B$  and the corresponding temperatures by
$T_{M}$, we observe from Eq. \eqref{UN2} that all the curves
merge into a single one, thereby demonstrating the scaling
behavior typical for HF metals \cite{shaginyan:2010}. As seen
from Fig. \ref{FIG7_1}, $\chi_N$ extracted from measurements on
$\rm Au_{51}Al_{34}Yb_{15}$ \cite{deguchi:2012} exhibits the
scaling behavior given by Eq. \eqref{UN2} and our calculations
agree well with the data over four orders of magnitude in
normalized temperature. Also seen is that $\chi_N$ agrees well
with the normalized $(C/T)_N$ extracted from measurements in
magnetic fields on $\rm YbRh_2Si_2$ \cite{oeschler:2008}. At
$T_N>1$ the susceptibility $\chi_N(T)\propto T^{-\beta}$, with
$\beta=1/2$. It is seen from Figs. \ref{FIG7_1} and \ref{FIG7},
panel (b), that the $T^{-\beta}$ regime with $\beta=1/2$ is
marked as NFL since contrary to the LFL case, the effective mass
depends strongly on temperature. { Additionally, it is observed
the robustness of the NFL phase against application of pressure
$P$ in zero magnetic field, in that the divergent $T$ and $B$
dependencies of $\chi$ are conserved \cite{deguchi:2012}. In
contrast to resilience of these divergences under pressure,
application of even a tiny magnetic field $B$ is sufficient to
suppress them, leading to Landau Fermi liquid (LFL) behavior at
low temperatures, as we have seen above. Thus, the paramagnetic
NFL phase of $\rm Au_{51}Al_{34}Yb_{15}$ takes place without
magnetic criticality, and not from quantum critical
fluctuations, strongly resembling the corresponding behavior
observed in $\rm \beta-YbAlB_4$
\cite{shaginyan:2013:A,shag:2016}.}

Again, we conclude that the QC $\rm Au_{51}Al_{34}Yb_{15}$
belongs to the family of HF compounds, and, thus, demonstrates
the new state of matter.

\section{One-dimensional quantum spin liquid and other possible realizations
of HF compounds}\label{TLL}

We have shown above that different HF compounds exhibit the
uniform and universal scaling in their thermodynamic, transport
and relaxation properties in the wide range of magnetic field,
temperature, number density, etc. Additionally to the already
known materials whose properties provide information on both the
existence of FCQPT and the new state of matter, there exists a
variety of other objects of great interest; these could be
studied and understood within the framework of the FC theory and
flat bands. Among such objects are neutron stars, atomic
clusters and fullerenes, ultra cold gases in traps, 1D quantum
spin systems, and the Universe itself
\cite{khod2001,shaginyan:2010,amusia:2015,shaginyan:2011:B,kono:2015}.

\begin{figure} [! ht]
\begin{center}
\vspace*{-0.5cm}
\includegraphics [width=0.45\textwidth]{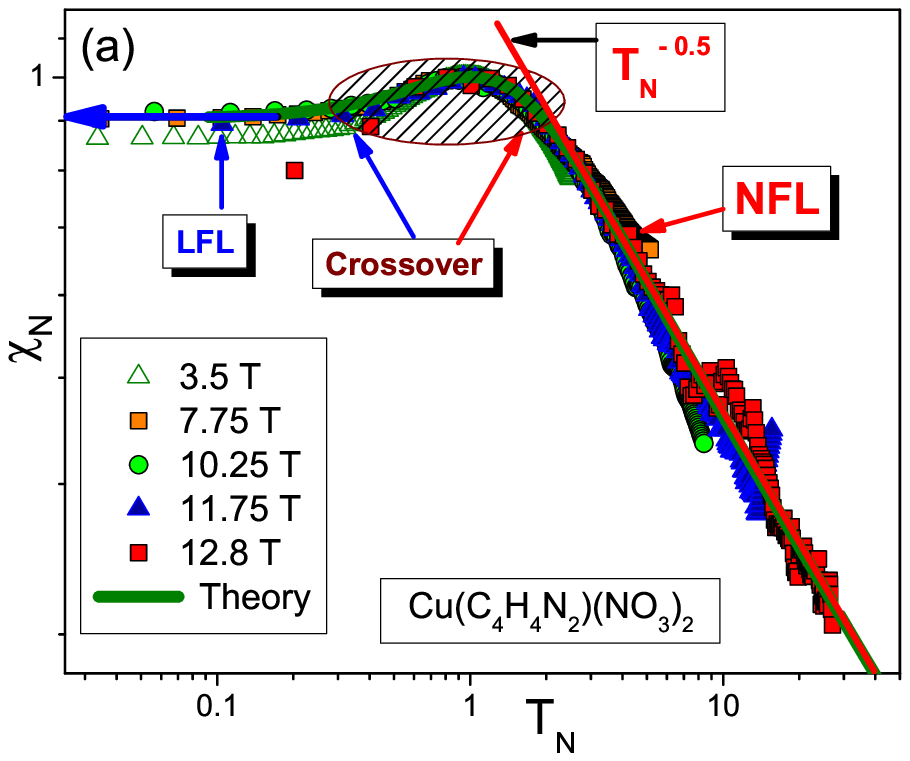}
\includegraphics [width=0.47\textwidth]{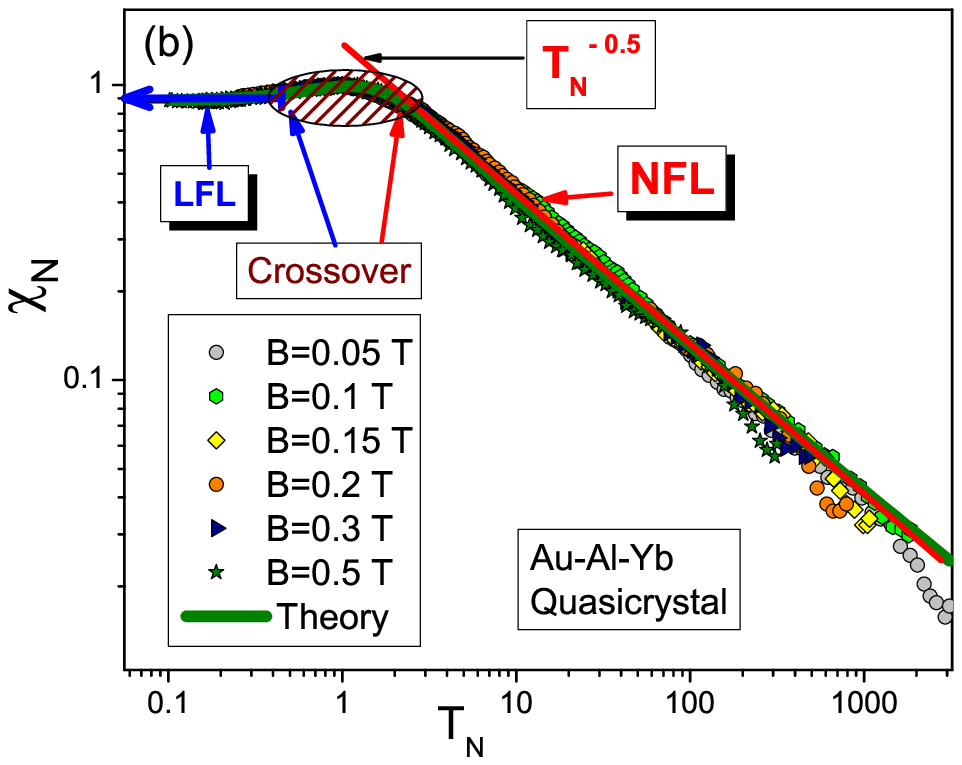}
\end{center}
\vspace*{-0.5cm} \caption{(color online) Universal scaling
behavior of both the insulator $\rm Cu(C_{4}H_4N_2)(NO_3)_2$
with the 1D quantum spin liquid and the quasicrystal $\rm
Au_{51}Al_{35}Yb_{14}$. Panel a: Temperature dependence on the
double logarithmic scale of the normalized magnetic
susceptibility $\chi_N$ versus normalized temperature $T_N$,
extracted from measurements of the magnetization on $\rm
Cu(C_{4}H_4N_2)(NO_3)_2$ at different magnetic fields
\cite{kono:2015} shown in the legend. The LFL and NFL regions
are shown by the solid arrows. The crossover region is depicted
by the shaded area. Panel b: Temperature dependence on the
double logarithmic scale of the magnetic susceptibility
$\chi_N$, versus normalized temperature $T_N$, extracted from
measurements of the magnetic susceptibility $\chi$ on the
quasicrystal $\rm Au_{51}Al_{35}Yb_{14}$ at different magnetic
fields \cite{deguchi:2012} shown in the legend. The LFL  and NFL
regions  are shown by the arrows. The crossover region is
depicted by the two arrows and dashed area. The solid line
marked by the NFL displays $M^*_N\propto T_N^{-0.5}$ behavior.
In both panels, our calculations are shown by the same solid
curve taken from Fig. \ref{FIG7_1}.}\label{FIG7}
\end{figure}
{ In 1D quantum spin liquid FCQPT plays a role of QCP taking
place at $H=H_s$. Here $H_s$ is the saturation magnetic field,
where QCP occurs, for near QCP taking place at $H=H_s$ and
$T=0$, the fermion spectrum becomes almost flat, and the fermion
(spinon) effective mass diverges, $M^*\propto M/p_{FH}\to
\infty$, due to kinematic mechanism, since the Fermi momentum
$p_{FH}\to 0$ of becoming empty subband
\cite{shaginyan:2016,shaginyan2016}. In other words, at $H=H_s$
both antiferromagnetic sublattices align in the field direction
i.e. the magnetic field fully polarizes the spins; at $H\geq
H_s$ the ground state is a gapped, field-induced ferromagnetic
state \cite{kono:2015}.} In case of weak repulsion between
spinons the divergence is associated with the onset of a
topological transition at finite value of $p_{FH}$ signaling
that $M^*(T)\propto T^{-1/2}$. Thus, the bare interaction of
spinons is weak \cite{kono:2015}. In that case the original
Tomonaga-Luttinger system
\cite{tomonaga:1950,luttinger:1963,haldane:1981,haldane:1980}
can exactly be mapped on a system of free spinons, which
low-temperature behavior in magnetic fields can be viewed as the
LFL one \cite{rozhkov:2005}. { On the other hand, the fermion
condensation associated with completely flat band does not occur
as it is absorbed by gapped, field-induced ferromagnetic phase
Fermi liquid at $H>H_s$ \cite{kono:2015}. As a result, this new
state at $H>H_s$ is protected by the gap rather than by the
Volovik topological number \cite{volovik:1991,volovik:2015}.
Nonetheless, at elevated temperatures the system bears
fingerprints of FC with the behavior very similar to that of HF
compounds with approximately flat bands
\cite{shaginyan:2016,shaginyan2016}, including the LFL and NFL
regimes. Note that recently a new state of matter, quasi-Fermi
liquid, has been introduced by Rozhkov \cite{rozhkov:2014} in
context of 1D Fermi liquid, while Lebed observed that
applicability of Fermi-liquid theory restores in
quasi-one-dimensional conductors \cite{lebed:2015}. This state
exhibits, however, some of the properties of ordinary Fermi
liquid accompanied by the NFL behavior.} Thus $\rm
Cu(C_{4}H_4N_2)(NO_3)_2$ (CuPzN) offers a unique possibility to
observe a new type of 1D quantum spin liquid whose thermodynamic
properties resemble that of HF compounds. Theory of 1D liquids
is still under construction and recent results show that the
liquids can exhibit LFL, non-Fermi liquid (NFL) and crossover
behavior \cite{rozhkov:2005,rozhkov:2014,lebed:2015}.

\begin{figure} [! ht]
\begin{center}
\includegraphics [width=0.50\textwidth]{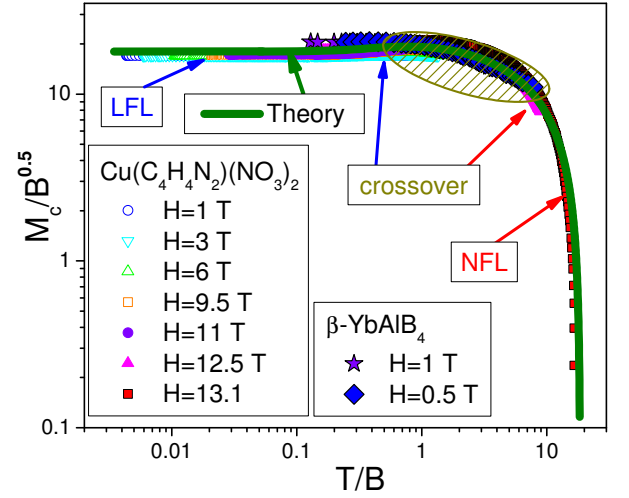}
\end{center}
\caption{(color online). The scaling dependence of $M_c/B^{0.5}$
versus $T/B$ for CuPzN and HF superconductor $\beta$ -
YbAlB$_4$. Here, $B=H_s-H$, see text for details. The
experimental data are taken from Refs.
\cite{kono:2015,s2011,mats:2012} and normalized making them
merge in the LFL region. The curves corresponding to different
magnetic fields $H$ are shown in the legends. The LFL, crossover
and NFL regions are shown. The theory is depicted by the solid
curve \cite{shaginyan:2016}.} \label{Fig1s}
\end{figure}
Recent measurements on the insulator CuPzN \cite{kono:2015}
allow us to analyze the properties of 1D quantum spin liquid
\cite{shaginyan:2016,shaginyan2016}. In Fig. \ref{FIG7}, panel
(a), we compare our calculated $\chi_N$ with that extracted from
measurements of the magnetization on single crystals of the
insulator \cite{kono:2015}. In order to understand the behavior
of $\chi_N$ displayed in the panel (a) of Fig. \ref{FIG7}, we
contrast $\chi_N$ extracted from measurements \cite{kono:2015}
with that obtained on the QC $\rm Au_{51}Al_{35}Yb_{14}$
displayed in the panel (b). It is seen from Fig. \ref{FIG7} that
both the normalized $\chi_N$ shown in the panels (a) and (b) are
in good agreement with each other. We now focus on the LFL, NFL
and the crossover LFL-NFL regions exhibited by the normalized
magnetic susceptibilities. To this end, we display in the panels
(a) and (b) of Fig. \ref{FIG7} the normalized $\chi_N$ on the
double logarithmic scale. As seen from Fig. \ref{FIG7}, the two
regions (the LFL and NFL regions) are clearly demarcated. And
good agreement of theory with the experimental data is realized.
The straight lines in Fig. \ref{FIG7}, panels (a) and (b),
capture both the LFL and NFL behaviors of $\chi_N\propto const$
and $\chi_N\propto T_N^{-1/2}$. These lines are in a good
agreement with the behavior of $M^*_N$ given by Eq. \eqref{UN2}
with $n=5/2$. Clearly, our calculations shown by the solid curve
in the panel (a) agree well with the measurements over three
orders of magnitude in the normalized temperature.

Taking into account that $M=\int\chi dH$ and Eqs. \eqref{UN2}
and \eqref{NORM}, we obtain that $(M_s-M)/\sqrt{H_s-H}$ as a
function of the variable $T/(H_s-H)$ exhibits scaling behavior.
Here $M_s$ is the saturation magnetization, taking place at
$H_s$. This result is in good agreement with the experimental
facts, as it is seen from Fig. \ref{Fig1s} that reports the plot
of the scaling behavior of the magnetization $M_c/B^{0.5}$. In
the case of CuPzN $M_c/B^{0.5}$ reads
$M_c/B^{0.5}=a+(M_s-M)/(H_s-H)^{0.5}$, being a function of
$T/B=T/(H_s-H)$, with $a$ is a constant added to a better
presentation of Fig. \ref{Fig1s}. In the case of $\rm \beta -
YbAlB_4$ magnetic field $H=B$ and $M_c$ is taken from Refs.
\cite{s2011,mats:2012}. It is seen from Fig. \ref{Fig1s}, that
the LFL behavior takes place at $T\ll B$, the crossover at
$T\sim B$, and the NFL one at $T\gg B$, as it is in the case of
the HF superconductor $\beta$ - YbAlB$_4$ \cite{shag2016}, see
Fig. \ref{Fig1s}, or the quasicrystal $\rm
Au_{51}Al_{34}Yb_{15}$, see Fig. \ref{FIG7}
\cite{deguchi:2012,shaginyan:2013:A}. We note that the observed
NFL behavior of of CuPzN is extremely sensitive to a magnetic
field, as it is in the case of both $\rm Au_{51}Al_{34}Yb_{15}$
and $\beta$ - YbAlB$_4$. It is worthy of note that the scaling
behavior of the thermodynamic properties of CuPzN has been
observed in Ref. \cite{jeong:2015}.

As a result, we conclude that 1D quantum spin liquid of the
insulator $\rm Cu(C_{4}H_4N_2)(NO_3)_2$ exhibits the typical
scaling behavior of its thermodynamic properties, and belongs to
the famous family of HF compounds, and therefore forms the new
state of matter \cite{shaginyan:2016,shaginyan2016}. Thus, we
expect the quantum physics of the 1D quantum spin liquid to be
universal, and to emerge regardless of the underlying
microscopic details of insulators holding the liquid.

\section{Summary}\label{SUM}

{We have shown both analytically and using arguments based
entirely on the experimental facts that the data collected on
very different strongly correlated Fermi systems named HF
compounds and represented by HF metals, quantum liquids like 2D
$\rm ^3He$, insulators with 2D, 3D quantum spin liquids and 1D
quantum spin liquids, and quasicrystals, have the universal
scaling behavior, observed in their thermodynamic, transport and
relaxation properties. Thus, despite of their drastic
microscopic diversity, strongly correlated Fermi systems exhibit
the uniform scaling behavior.} The quantum physics of different
strongly correlated Fermi systems is universal and emerges
regardless of their underlying microscopic details. The observed
behavior resembles the uniform collective behavior exhibited by
such different states of matter as the gaseous, liquid and solid
states of matter. Therefore, this uniform behavior, formed by
flat bands, allows us to view it as the main manifestation of
the new state of matter exhibited by HF compounds.

\section{acknowledgement}

VRS is supported by the Russian Science Foundation, Grant No.
14-22-00281. AZM thanks the US DOE, Division of Chemical
Sciences, Office of Energy Research, and ARO for research
support. PKG is partly supported by RFBR \# 14-02-00044. VAK
thanks the McDonnell Center for the Space Sciences for support.

\end{document}